\documentclass[
twocolumn,
reprint,
amsmath,amssymb,aps,prb,superscriptaddress,floatfix,showpacs]{revtex4-2}

\usepackage[T1]{fontenc}            
\usepackage[utf8]{inputenc}
\usepackage{bm}
\usepackage{xspace}%
\usepackage{graphicx}
\usepackage{dcolumn}
\usepackage{bm}
\usepackage{tabularx}
\usepackage[rflt]{floatflt}
\usepackage{float} 
\usepackage{tikz}
\usepackage{url}
\usepackage{amsmath}
\usepackage{makeidx,amssymb,amsmath,amsthm,graphicx}
\usepackage[toc,page]{appendix}
\usepackage{comment}
\usepackage[
left=1.5cm,
right=1.5cm,
top=1.50cm,
bottom=1.50cm
]{geometry}

\usepackage{xcolor}

\usepackage{babel}
\newcommand{\pcm}{\ensuremath{\,\mathrm{cm}^{-1}}\xspace}
\newcommand{\pcmc}{\ensuremath{\,\mathrm{cm}^{-3}}\xspace}
\newcommand{\eVolt}{\ensuremath{\,\mathrm{eV}}\xspace}
\newcommand{\pOhmpcm}{\ensuremath{\,\mathrm{\Omega}^{-1}\mathrm{cm}^{-1}}\xspace}
\newcommand{\EuZnP}{EuZn$_2$P$_2$\xspace}
\newcommand{\EuCdP}{EuCd$_2$P$_2$\xspace}
\newcommand{\EuCdAs}{EuCd$_2$As$_2$\xspace}

\makeatother
\begin{document}
	
	
	\title{Colossal magnetoresistance in EuZn$_2$P$_2$ and its electronic and magnetic structure}
	\author{Sarah~Krebber}
	\affiliation{%
		Physikalisches Institut, Goethe-Universit\"at Frankfurt, Max-von-Laue Stasse 1, 60438 Frankfurt am Main, Germany
	}%
	\author{Marvin~Kopp}
	\affiliation{%
		Physikalisches Institut, Goethe-Universit\"at Frankfurt, Max-von-Laue Stasse 1, 60438 Frankfurt am Main, Germany
	}%
	\author{Charu~Garg}
	\affiliation{%
		Physikalisches Institut, Goethe-Universit\"at Frankfurt, Max-von-Laue Stasse 1, 60438 Frankfurt am Main, Germany
	}%
	\author{Kurt~Kummer}
	\affiliation{European Synchrotron Radiation Facility (ESRF), 38043 Grenoble, France}

	\author{J\"org~Sichelschmidt}
	\affiliation{Max Planck Institute for Chemical Physics of Solids, N\"othnitzer Stra\ss e 40, 01187 Dresden, Germany}
	\author{Susanne~Schulz}
	\affiliation{Institut f\"ur Festk\"orper- und Materialphysik, Technische Universit\"at Dresden, D-01062 Dresden, Germany}
 
	\author{Georg~Poelchen}
	  \affiliation{European Synchrotron Radiation Facility (ESRF), 38043 Grenoble, France}
	\affiliation{Institut f\"ur Festk\"orper- und Materialphysik, Technische Universit\"at Dresden, D-01062 Dresden, Germany}
    \affiliation{Max Planck Institute for Chemical Physics of Solids, N\"othnitzer Stra\ss e 40, 01187 Dresden, Germany}
 
	\author{Max~Mende}
	\affiliation{Institut f\"ur Festk\"orper- und Materialphysik, Technische Universit\"at Dresden, D-01062 Dresden, Germany}
 
    \author{Alexander~V.~Virovets}
	\affiliation{%
		Institute of Inorganic Chemistry, Goethe-Universit\"at Frankfurt, Max-von-Laue Stasse 7, 60438 Frankfurt am Main, Germany
	}%
	\author{Konstantin~Warawa}
	\affiliation{%
		Physikalisches Institut, Goethe-Universit\"at Frankfurt, Max-von-Laue Stasse 1, 
		60438 Frankfurt am Main, Germany
	}%
	\author{Mark~D.~Thomson}
	\affiliation{%
		Physikalisches Institut, Goethe-Universit\"at Frankfurt, Max-von-Laue Stasse 1, 
		60438 Frankfurt am Main, Germany
	}%
	\author{Artem~V.~Tarasov}
	\affiliation{Donostia International Physics Center (DIPC), 20018 Donostia-San Sebastián, Spain}
	\author{Dmitry~Yu.~Usachov}
	\affiliation{Donostia International Physics Center (DIPC), 20018 Donostia-San Sebastián, Spain}
	\author{Denis~V.~Vyalikh}
	\affiliation{Donostia International Physics Center (DIPC), 20018 Donostia-San Sebastián, Spain}
	\affiliation{IKERBASQUE, Basque Foundation for Science, 48013 Bilbao, Spain}
	\author{Hartmut~G.~Roskos}
	\affiliation{%
		Physikalisches Institut, Goethe-Universit\"at Frankfurt, Max-von-Laue Stasse 1, 
		60438 Frankfurt am Main, Germany
	}%
	\author{Jens~M\"uller}
	\affiliation{%
		Physikalisches Institut, Goethe-Universit\"at Frankfurt, Max-von-Laue Stasse 1, 
		60438 Frankfurt am Main, Germany
	}%
	\author{Cornelius~Krellner}
	\affiliation{%
		Physikalisches Institut, Goethe-Universit\"at Frankfurt, Max-von-Laue Stasse 1, 60438 Frankfurt am Main, Germany
	}%
	\author{Kristin~Kliemt}
	\email[Corresponding author:]{kliemt@physik.uni-frankfurt.de}
	\affiliation{%
		Physikalisches Institut, Goethe-Universit\"at Frankfurt, Max-von-Laue Stasse 1, 60438 Frankfurt am Main, Germany
	}%

	\date{\today}
	\begin{abstract}
We investigate single crystals of the trigonal antiferromagnet EuZn$_2$P$_2$ ($P\overline{3}m1$) by means of electrical transport, magnetization measurements, X-ray magnetic scattering, optical reflectivity, angle-resolved photoemission spectroscopy (ARPES) and \textit{ab initio} band structure calculations (DFT+U). 
We find that the electrical resistivity of EuZn$_2$P$_2$ increases strongly upon cooling and can be suppressed in magnetic fields by several orders of magnitude (CMR effect).  
Resonant magnetic scattering reveals a magnetic ordering vector of $q = (0\, 0\, \frac{1}{2})$, corresponding to an $A$-type antiferromagnetic (AFM) order, below $T_{\rm N}$ = 23.7\,K. We find that the moments are canted out of the $a-a$ plane by an angle of about $40^{\circ}\pm 10^{\circ}$ degrees and aligned along the [100] direction in the $a-a$ plane.
We observe nearly isotropic magnetization behavior for low fields and low temperatures which is consistent with the magnetic scattering results. 
The magnetization measurements show a deviation from the Curie-Weiss behavior below $\approx 150\,\rm K$, the temperature below which also the field dependence of the material's resistivity starts to increase. An analysis of the infrared reflectivity spectrum at $T=295\,\rm K$ allows us to resolve the main phonon bands and intra-/inter-band transitions, and estimate indirect and direct band gaps of $E_i^{\mathrm{opt}}=0.09\,\rm{eV}$ and $E_d^{\mathrm{opt}}=0.33\,\rm{eV}$, respectively, which are in good agreement with the theoretically predicted ones.
The experimental band structure obtained by ARPES is nearly $T$-independent above and below $T_{\rm N}$. The comparison of the theoretical and experimental data shows a weak intermixing of the Eu 4$f$ states close to the $\Gamma$ point with the bands formed by the phosphorous 3$p$ orbitals leading to an induction of a small magnetic moment at the P sites.
\end{abstract}
	
	\keywords{Growth from high-temperature solutions, Single crystal growth, Rare earth compounds, Eu compounds, CMR effect}
	\maketitle

	\section{\label{sec:level1}Introduction}

      	\begin{figure*}[ht]
		\begin{center} 
					\includegraphics[width=0.98\textwidth]{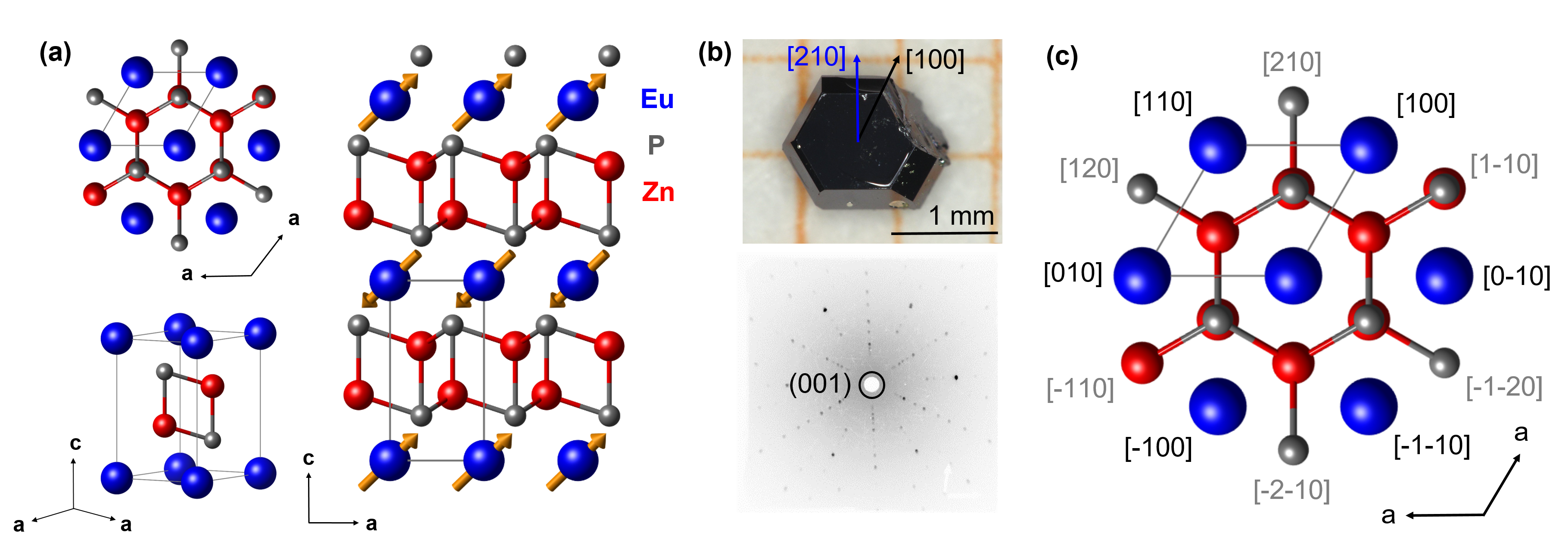}
		\end{center}
		\caption{(a) The trigonal CaAl$_2$Si$_2$-type crystal structure of EuZn$_2$P$_2$ with the space group $P\overline{3}m1$ (No. 164). The orange arrows indicate the magnetic structure of the Eu atoms below $T_{\rm N}$. (b) Single crystal and Laue pattern of EuZn$_2$P$_2$. (c) Defining the in-plane axes of the trigonal crystal structure of EuZn$_2$P$_2$ in the direct lattice.}
		\label{Crystal_Structure}
	\end{figure*}

	Today, more and more trigonal Eu-based compounds of the EuT$_2$X$_2$ (T = Cd, Zn; X = P, Sb, As) family with  CaAl$_2$Si$_2$-type structure in the space group $P\overline{3}m1$ (164) come into focus as magnetic topologically non-trivial materials. In recent studies, antiferromagnetic (AFM) EuCd$_2$As$_2$, $T_{\rm N}=9.5\,\rm K$ \cite{Schellenberg2011} with  $A$-type AFM structure and in-plane moments \cite{Rahn2018}, has attracted considerable attention due to the observation of strong coupling between charge transport and magnetism \cite{Wang2016} as well as the emergence of topological phases \cite{Ma2019}. 
	In the paramagnetic phase of EuCd$_2$As$_2$, a "spin-fluctuation-induced" Weyl semimetal state \cite{Ma2019} and its tunability by pressure \cite{Gati2021} was discovered. Using first-principles band structure calculations it was predicted that EuCd$_2$As$_2$ in a ferromagnetic configuration (with out-of-plane moments) can host a single pair of Weyl points due to the half-metallic nature of the material \cite{Wang2019}. This has recently been confirmed experimentally \cite{Soh2019} making EuCd$_2$As$_2$ a model semimetal suitable for fundamental tests of Weyl physics. Furthermore, the related AFM material EuCd$_2$P$_2$, $T_{\rm N}=11.6\,\rm K$ \cite{Schellenberg2011}, shows a  colossal magnetoresistance (CMR) effect \cite{Wang2021}. Recently, Sunko \textit{et al.} discovered the onset of ferromagnetic order above  $T_{\rm N}$ in the temperature range of a pronounced resistivity peak \cite{Sunko2022}.
	Using Monte-Carlo simulations, Heinrich \textit{et al.} explored the magnetic phases predicted by a microscopic classical magnetic model for EuCd$_2$P$_2$, providing qualitative numerical evidence that the CMR effect is related to a magnetic Berezinskii-Kosterlitz-Thouless (BKT) transition \cite{Heinrich2022}.\\
	This variety of interesting topological and magnetic phenomena in the trigonal EuT$_2$X$_2$ systems also led to the first investigation of a related compound, namely EuZn$_2$P$_2$ \cite{Berry2022}. Insulating behavior was found, when measured down to $140\,\rm K$ and $A$-type AFM order with magnetic Eu$^{2+}$ moments aligned in the $a-a$ plane has been inferred from combined magnetic measurements and density functional theory (DFT) calculations \cite{Berry2022}. 
    Anticipating a strong coupling of the charge transport with the system's magnetization, we investigate the magnetic and electronic properties of EuZn$_2$P$_2$ in detail down to low temperatures. 
    We performed a comprehensive study of the magnetic ground state of the material combining magnetization measurements and X-ray magnetic scattering. In transport measurements, we observe a CMR effect at low temperatures in EuZn$_2$P$_2$, a system not being a mixed-valent perovskite manganite and not a ferromagnet \cite{Hwang1995, Solovyev1996, Salamon2001}.
    Similar to the CMR effect in EuCd$_2$P$_2$, we find that in EuZn$_2$P$_2$ the resistivity becomes strongly suppressed in a magnetic field upon approaching the AFM ordering temperature with the onset of the large negative magnetoresistance (\textit{MR}) occurring at temperatures as high as six times $T_{\rm N}$.     
    Furthermore, we present ARPES data together with the modeling of the band structure using DFT+U calculations as well as optical reflectivity data  allowing us to determine the indirect and direct band gaps.

	
	\section{Experimental details}
	\vspace*{-\baselineskip}
	\subsection{Crystal growth}
    \vspace*{-\baselineskip}
	Single crystals of trigonal EuZn$_2$P$_2$, Fig.~\ref{Crystal_Structure}, were grown from an external Sn flux by using ingots of europium (99.99\,\%, Evochem), red phosphorous (99.9999\,\%, Chempur), tin (99.999\,\%, Evochem) and teardrops of zinc (99.9999\,\%, Chempur). All elements were cut into small pieces and mixed together with a molar ratio of Eu:Zn:P:Sn = 1:2:2:20. This stoichiometry was initially utilized in analogy to the recipe used in \cite{Wang2021} for EuCd$_2$P$_2$. All reactants and products were handled under an inert Ar atmosphere inside a glove box. The starting materials were put in a graphite crucible inside an evacuated quartz ampule. The ampule was then loaded into a box furnace (Thermconcept), heated up to $450\,^\circ\text{C}$ and held for $5\,\rm h$.
	This ensures that the phosphorous slowly reacts with the other materials. Afterwards, the temperature was raised to $1100\,^\circ\text{C}$ and held there for a few hours in order to homogenize the melt. The temperature was then slowly lowered to $600\,^\circ\text{C}$ with a rate of $2\,\text{K}$/h, where the liquid flux was then removed by centrifuging. Several optimization steps with respect to stoichiometry and temperature profile were carried out in order to obtain large and high-quality single crystals. Hexagonal-shaped single crystals with an average size of $2\,\rm mm\times 2\,\rm mm$ were obtained, see Fig.~\ref{Crystal_Structure}(b). Remaining tin on the surfaces of the crystals was removed mechanically.

	\subsection{Structural, chemical and magnetic characterization}
    \vspace*{-\baselineskip}
 
	X-ray powder diffraction (PXRD) was used to confirm the trigonal space group $P\overline{3}\text{m}1$ (No. 164) on powdered single crystals, Fig.~\ref{Crystal_Structure}. The refined lattice parameters $a=b=4.0871$\,\AA\, and $c=7.0066$\,\AA\, are in good agreement with the literature \cite{Frik99, Berry2022}. The diffraction patterns were recorded on a diffractometer with Bragg-Brentano geometry and copper K$_\alpha$ radiation.
 
    Additionally, we performed single crystal X-ray diffraction on a small platelet selected from a EuZn$_2$P$_2$ sample. This data were collected at 212\,K on STOE IPDS II two-circle diffractometer equipped with the Genix 3D HS microfocus Mo K$\alpha$ X-ray source ($\lambda$ = 0.71073\,\AA). The finalization of the data, including empirical absorption corrections, was done using the CrysAlisPro software (Rigaku Oxford Diffraction, 2022). The initial structural model was taken from the literature \cite{Berry2022}. The structure was refined in the anisotropic approximation against ${|F|}^2$ with full-matrix least-squares techniques using the program SHELXL-2018/3 \cite{shelx2015}. Crystallographic data and parameters of the diffraction experiments are given in Table \ref{SingleCrystalDiff}. The CIF file is deposited in Cambridge Crystallographic Data Center under the code CSD 2262937 \cite{CD2262937}. The results show that the crystal structure is fully consistent with the previously reported data by \cite{Berry2022}. At that, the refinement of the site occupancy factor of the phosphorous atom gives no signs for any off-stoichiometry and the diffraction pattern can be fully indexed in the hexagonal unit cell without any traces of modulation. Therefore, we can exclude pronounced structural defects in our crystals of EuZn$_2$P$_2$.

    \vspace*{-\baselineskip}
        \begin{center}
    \begin{table}[!ht]
    \caption{Crystallographic data and refinement results for EuZn$_2$P$_2$ from Single Crystal X-ray diffraction \cite{CD2262937}.}
    \label{SingleCrystalDiff}
    \begin{tabularx}{\linewidth}{ll} 
    \multicolumn{2}{l}{}\\
    \hline
    \hline
     Material & EuZn$_2$P$_2$\\ 
    \hline
    M$_r$ & 344.64 \\ 
    Crystal system  &Trigonal \\ 
    Space group & P$\overline{3}$m1\,(\text{No.}\,164) \\
    Temperature (K) & 212 \\
    Lattice parameters  & \\
    a (\,\AA\,) & 4.08582(10) \\
    c (\,\AA\,) & 7.0041(3) \\
    V ({\,\AA\,}$^3$) & 101.26(1) \\
    Z & 1 \\
    \textit{F}(000) & 153 \\
    \textit{D$_x$} (Mg m$^{-3}$) & 5.652 \\
    Radiation type & Mo \textit{K$\alpha$} \\
    $\mu$ (mm$^{-1}$) & 27.59 \\
    Crystal size (mm) & 0.07 × 0.06 × 0.03 \\
    Crystal shape, colour & Platelet, black \\
    \textit{R$_{\text{int}}$} & 0.081 \\
    2$\Theta_{max}$ (°)	 & 67.2 \\
    \textit{$R(F)^a, wR(F^2)^b, GooF^c$ \qquad \qquad} & 0.016, 0.037, 1.16 \\
    No. of parameters&  10 \\
    $\Delta \rho_{max}$,  $\Delta \rho_{min}$ ($ e \text{\,\AA\,}^{-3}$)& 1.11, -1.32 \\
    \hline
    \multicolumn{2}{l}{}\\
    \multicolumn{2}{l}{$^a R(F) = \sum||F_o|-|F_c||/\sum|F_o|$ for \textit{F$^2$}$ >2\sigma (F^2)$} \\
    \multicolumn{2}{l}{$^bwR(F^2) = [\sum w({F_o}^2-{F_c}^2)^2/\sum w({F_o}^2)]^{1/2}$} \\
    \multicolumn{2}{l}{$^cGooF = [(\sum w({F_o}2{F_c}^2)^2)/(N_{\text{ref}}-N_{\text{param}})]^{1/2}$} \\
    \multicolumn{2}{l}{for all reflections}\\
    \end{tabularx}
    \end{table}
    \end{center}
    \vspace*{-\baselineskip}
     
     The orientation of the single crystals was determined using the Laue method and the spectra were simulated using QLaue software as shown in the appendix~\ref{EDX_Laue}. Fig.~\ref{Crystal_Structure}(b) shows the recorded Laue pattern with the beam parallel to the $c$-axis. The sharp spots in the pattern suggest a high crystallinity of the samples and additionally the sixfold symmetry of the samples can be seen. The chemical composition of the EuZn$_2$P$_2$ single crystals was confirmed with energy-dispersive X-ray spectroscopy (EDX). A Quantum Design physical property measurement system (PPMS) was used to investigate the magnetic properties and the heat capacity of EuZn$_2$P$_2$. The heat capacity measured on our single crystals, Fig.~\ref{HC_EuZn2P2}, confirms the occurrence of the AFM transition at $T_{\rm N}=23.7\,\rm K$ and is consistent with the previously published data in Ref.~\cite{Berry2022}. The grown crystals are hexagonal shaped platelets, where the direction perpendicular to the platelet is along the [001]-direction. The natural crystal edges are perpendicular to the $[210]$-direction in real space and the corners of the crystal are along the $[100]$-direction, Fig.~\ref{Crystal_Structure}(c). This notation will be used to describe the direction of the applied magnetic field in the magnetization measurements. Note that in the trigonal space group 164,
     the angle between the basis vector $a$ in real space and the basis vector $a^{*}$ in reciprocal space is 30$^{\circ}$. 

	\subsection{Electrical transport}\label{sec:rho_details}
    \vspace*{-\baselineskip}
	Electrical transport measurements were carried out on polished EuZn$_2$P$_2$ samples with thermally evaporated chromium/gold layers (7\,{\rm nm}/200\,{\rm nm} thickness) contacted by conducting silver paste in order to ensure ohmic behavior of the contacts, which were checked by measuring the $I$-$V$ characteristics at different temperatures. The evaporation process also allows for a well-defined contact geometry on the $a$-$a$ surface in a desired four-point configuration. We used a low-frequency lock-in technique to measure the resistance $R = V/I$ in the ohmic regime and calculated the resistivity from the contact geometry in the $a$-$a$ plane and the dimensions along $c$. Furthermore, we found that without polishing the crystals and with contacts painted with silver epoxy directly on the surface of the crystals we observe nonlinear I-V curves and the temperature dependence down to 150\,K was similar to what was observed in Ref. \cite{Berry2022}.

	\subsection{Resonant magnetic X-ray diffraction}
	\vspace*{-\baselineskip}
	Resonant magnetic diffraction measurements were performed at the ID32 soft X-ray beamline of the ESRF \cite{brookes2018-nima}. We used the four-circle diffractometer with a photodiode detector in horizontal diffraction geometry mounted in the RIXS endstation. An Apple-II type undulator operated on its first harmonic delivered the $\pi -$ and $\sigma -$ polarized incident beam. The photon energy was set to slightly below the white line in the Eu$^{2+}$ M$_5$ absorption spectrum. The beam spot size at the sample was $10\times 2\,$\textmu m$^2$. We cleaved the sample inside the vacuum chamber in order to obtain a surface with larger terraces compared to the as-grown surface. In this way, we were able to obtain magnetic domains larger than the beam spot size on which we could measure the azimuth dependence of the $\sigma/\pi$ intensity ratio of the magnetic Bragg peak.

	\subsection{DFT+U calculations}
    \vspace*{-\baselineskip}
	The electronic structure calculations were performed using the OpenMX code which provides a fully relativistic DFT implementation with localized pseudoatomic orbitals \cite{ozaki2003variationally,ozaki2004numerical, ozaki2005efficient} and norm-conserving pseudopotentials \cite{troullier1991efficient}. The exchange-correlation energy in local spin density approximation was employed \cite{perdew1992}. The accuracy of the real-space numerical integration was specified by the cutoff energy of 200\,Ry, and the total energy convergence criterion was $10^{-7}$\,eV. The $\mathbf k$-mesh for Brillouin zones were specified as $12 \times 12 \times 6$. The basis functions were taken as P7.0-s$^2$p$^2$d$^1$f$^1$, Zn6.0H-s$^3$p$^2$d$^1$, Eu8.0-s$^3$p$^2$d$^2$f$^1$ (the pseudopotential cutoff radius is followed by a basis set specification). Experimental unit cell parameters were used, while the atomic positions in the unit cell were relaxed until the forces on each atom were less than 0.0005\,Hartree/Bohr ($\approx$ 10$^{-2}$\,eV/\AA). To account for the strong correlations of the $f$ electrons, the Eu~4$f$ states were treated within the DFT+U approach~\cite{han2006n} using the Dudarev scheme \cite{dudarev1998electron}. The Hubbard $U$ parameter was varied and the corresponding band structure was found to best match the experimental ARPES pattern for $U=1.8$~eV.
		
	\subsection{Angle-resolved photoemission spectroscopy}
    \vspace*{-\baselineskip}
	The presented ARPES measurements were performed at the UARPES beamline of the SOLARIS National Synchrotron Radiation Centre with a photon energy of 137\,eV and linear horizontal light polarization at a temperature of around 30\,K. The endstation is equipped with a Scienta DA30L analyzer. The EuZn$_2$P$_2$ single crystals were cleaved \emph{in situ} at a pressure better than 10$^{-10}$\,mbar to ensure an atomically clean surface. Due to the high resistivity of EuZn$_2$P$_2$ at low temperatures, charging of the sample resulted in an energy shift of the spectrum which was corrected by high-temperature reference measurements performed at the BLOCH beamline of the MAX-IV synchrotron radiation facility. 
 
	\subsection{Infrared spectroscopy}\label{sec:ir_details}
    \vspace*{-\baselineskip}
    The ground-state infrared excitations were investigated at $T=295$\,K via the near-normal-incidence reflectivity spectrum $R(\omega)$ of the (001)-surface (polished with 0.3\,\textmu m grain size) using a rapid-scan Michelson-type Fourier-transform spectrometer, covering the spectral range from 10\,meV to 1\,eV. %
    To obtain absolute values of $R$, the sample was coated \emph{in situ} with gold and then used for measuring the reference spectrum. %
    To avoid any artifacts that might arise from extrapolation outside the measured spectral range, as needed for a Kramers-Kronig analysis to estimate the complex conductivity $\sigma(\omega)=\sigma_1(\omega)+i\sigma_2(\omega)$ \cite{dresselbook}, we fit the reflectivity spectrum $R(\omega)$ directly with a multi-band model, comprising a standard Drude response $\sigma_D(\omega)=\sigma_{0D}/(1+i\omega\tau_D)$
    \cite{dresselbook}, two phonon bands, and two bands at higher energy to account for both, indirect and direct interband transitions (note we employ the $e^{+i\omega t}$ complex sign convention here).
    As discussed in Sec.~\ref{sec:infrared} below, the phonon signatures indicate that their bands deviate from a basic Lorentzian model, 
    $\sigma_L(\omega)=i\omega\Gamma\sigma_{0L}(\omega_0^2-\omega^2+i\Gamma\omega)^{-1}$ (which did not allow a detailed fit of $R(\omega)$ around each phonon), and hence we employ the more general Fano lineshape $\sigma_F(\omega)=\sigma_L(\omega)(q+i)^2/(1+q^2)$ \cite{sedlmeier12}, where $1/q\neq 0$ corresponds to a finite coupling between the phonon response with the spectrally overlapping Drude ``continuum'' response, resulting in an oscillatory distortion of the phonon bands (essentially mixing the real and imaginary parts of $\sigma_L$).
    For the interband transitions, we employ a Tauc-Lorentz band model with explicit onset bandgap energies $E_{g}=\hbar \omega_{g}$ ($g=i,d$), where the real part of the conductivity is given by  
    $\sigma_{g1}(\omega)=\sigma_{L1}(\omega)\omega^{-1}(\omega-\omega_g)^2\Theta(\omega-\omega_g)$, and one can obtain $\sigma_{g2}(\omega)$ also in closed form \cite{jellison96,jellison1996erratum}.
    Note that the Lorentzian factor in $\sigma_{g}(\omega)$ then provides a phenomenological model of the interband joint-density of states, where $\sigma_{g1}$ has a peak near $\omega_0$ and decays for $\omega-\omega_0\gg \Gamma$.
    The fitting procedure calculates the complex field reflection coefficient $\hat{r}(\omega)=\tfrac{1-\sqrt{\varepsilon_r}}{1+\sqrt{\varepsilon_r}}$ with $\varepsilon_r=\varepsilon_{r\infty}+(i\varepsilon_0\omega)^{-1}\sigma(\omega)$ (where $\sigma$ is the sum over all model bands) and minimizes the least-squares error of $\hat{R}(\omega)=|\hat{r}(\omega)|^2$, where $\varepsilon_{r\infty}$ is also treated as a fit parameter.

    \subsection{Carrier density calculations}\label{sec:density}
    \vspace*{-\baselineskip}
    We estimate the carrier density at T=300\,K from the fitted Drude conductivity parameters, and compare this to the value predicted on the basis of thermally excited, intrinsic carriers, using the effective masses based on the mass tensor elements (about the $M$-points in the conduction band and $\Gamma$-point for the valence band), and predicted an indirect band gap $E_i=0.2$~eV.
    The mass elements are given by $m_e^{j}/m_0=0.898, 0.134, 0.818$ along the directions $j=M\Gamma, MK, ML$, respectively, and $m_{h}/m_0=0.196$ ($m_0$ the free electron mass).
    For the Drude approach, we use the relation $\sigma_{0D}=Ne^2\tau_D/m^{aa}$ to solve for $N$, where $m^{aa}$ is the total effective conductivity mass $1/m^{aa}=1/m_{e}^{aa}+1/m_h$ \cite{lundstrombook}.  For the electrons at the $M$-point (Fig.~\ref{EuZn_DFT}), we take the effective electron conductivity mass for motion in the $a-a$ plane to be given by $1/m_{e}^{aa} =\tfrac{1}{2}(1/m_{e}^{M\Gamma} + 1/m_{e}^{MK})$, which yields $m_{e}^{aa}=0.233\cdot m_0$, and hence $m^{aa}=0.106\cdot m_0$. For the prediction based on thermal excitation of intrinsic carriers, one has $N=\sqrt{N_C N_V} \cdot e^{-E_i/2kT}$, where $N_C=2g_M(2\pi m_e^{DOS}k T/h^2)^{3/2}$ and $N_V=2(2\pi m_h k T/h^2)^{3/2}$ are the conduction- and valence-band density of states \cite{lundstrombook}, respectively, $g_M=3$ is the valley degeneracy of the 6 boundary-edge $M$ points where $m_e^{DOS}=\sqrt[3]{m_e^{M\Gamma}\cdot m_e^{MK}\cdot m_e^{ML}}=0.462\cdot m_0$.
	
\section{Results and discussion}
 \vspace*{-\baselineskip}
\subsection{Electronic transport measurements}\label{sec:transport}
\vspace*{-\baselineskip}
\begin{figure}[b]
\centering
\includegraphics[width=0.47\textwidth]{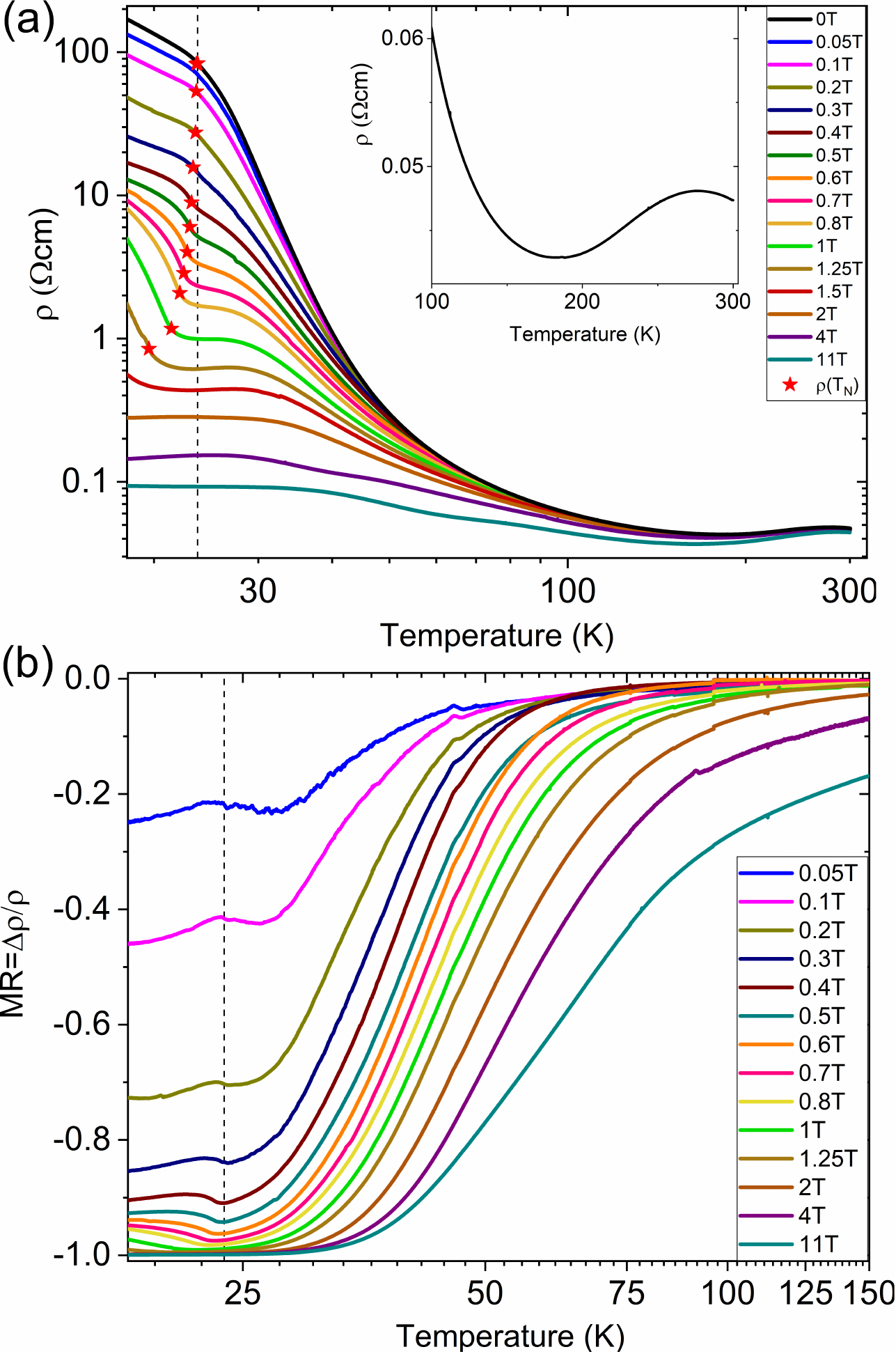}
\caption{(a) In-plane resistivity as a function of temperature for various magnetic fields $\mu_0 H$ applied out-of-plane along the $c$-axis. The inset shows a region of decreasing resistivity in zero field upon cooling down to $T\approx 180\,$K, where a minimum in the resistivity is observed. Upon further cooling, semiconducting behavior is observed with a change of slope/flattening below $T_{\rm N} = 23.7$\,K. The resistivity becomes strongly suppressed in increasing magnetic fields. The red stars mark the magnetic field-dependent transition temperatures $T_{\rm N}$ derived from magnetization measurements (see Figs.~\ref{MvT_MvH_001}(a) and \ref{PD_001_100_210}(a) in the supplemental information).   
(b) Magnetoresistance \textit{MR}=$[\rho(H)-\rho(H=0)]/\rho(H=0)$ calculated from the data in (a) for temperatures below $150\,$K. A large negative \textit{MR} of $-99.4\,$\% at $\mu_0H=4\,$T ($-99.9$\,\% for 11\,T) saturates at the transition temperature $T_{\rm N}$ in zero magnetic field marked by dotted lines in both graphs.} 
\label{rho_MR}%
\end{figure}%

Electrical transport measurements were carried out on polished EuZn$_2$P$_2$ samples. The temperature-dependent resistivity is shown in Fig.\,\ref{rho_MR}(a) in a double logarithmic plot for various magnetic fields up to $\mu_0H=11\,$T applied along the $c$-axis. In contrast to a previous report \cite{Berry2022}, we do not observe an Arrhenius behavior at zero field below room temperature but rather a small maximum upon cooling followed by a positive temperature coefficient ${\rm d}\rho/{\rm d}T$ above a minimum around $T\approx 180\,$K, see the inset of Fig.\,\ref{rho_MR}(a). Such behavior at elevated temperatures may be attributed to in-gap states which donate additional carriers and freeze out upon lowering the temperature which is in agreement with findings from infrared spectroscopy, see Sec.~\ref{sec:infrared}. Below about 180\,K a semiconducting behavior is resumed with a strongly increasing resistance down to a few Kelvin above $T_{\rm N} = 23.7$\,K, where the slope changes and a flattening is observed below the magnetic transition. It is important to note that the $I$-$V$ curves become increasingly nonlinear for temperatures below $T_{\rm N}$. Therefore, we refrain from discussing the resistivity behavior below $T=18\,$K. For increasing magnetic fields, the resistivity is suppressed at all temperatures as discussed in the following.\\  
Fig.\,\ref{rho_MR}(b) highlights the resistivity behavior in finite magnetic fields applied along the $c$-axis plotted as the magnetoresistance ratio \textit{MR} $= [\rho(H)-\rho(H=0)]/\rho(H=0)$. Even for small fields, a considerable negative magnetoresistance of order a few \% is observed already at temperatures as high as 150\,K. The \textit{MR} then becomes increasingly stronger in increasing magnetic fields upon lowering the temperature down to $T_{\rm N}$ (see below).
Such a strong negative \textit{MR}, reaching saturation values of $-99.4$\,\% at $\mu_0H = 4$\,T and $-99.9$\,\% at $11$\,T, is rather similar to the colossal negative magnetoresistance (CMR) observed in rare earth chalcogenides, Cr-based spinel systems, europium hexaboride or mixed-valence perovskite manganites, see e.g. \cite{Methfessel1968, Ramirez1997, Tokura1999, Dagotto2001, Tokura2006, Lin2016}.
Whereas most of these materials are ferromagnets, the CMR effect in antiferromagnetically ordered materials is less studied and recently has attracted considerable interest \cite{Rosa2020, AleCrivillero2023, Wang2021, Sunko2022}.

Another striking observation is a pronounced change of slope of the resistivity at $T_{\rm N}$ which shifts with increasing field to lower temperatures following $T_{\rm{N}}(H)$, see Fig.\,\ref{rho_MR}(a). In addition, the flattening of the resistivity below a kink at $T_{\rm N}$ at small magnetic fields develops into a plateau-like behavior for fields $\mu_0H \gtrsim 0.5$\,T the onset of which shifts to higher temperatures with increasing field. Likewise, the crossover temperature from weak to strong negative \textit{MR} behavior characterized by the inflection point of the \textit{MR} vs.\ temperature curves shifts to higher temperatures with increasing fields with an initial strong increase, i.e.\ for small fields up to 0.5\,T (see Fig.\,\ref{SI_inflection-point} in the appendix) followed by a more moderate increase for higher fields. Remarkably, for fields of order 1\,T the relative \textit{MR} amounts to $-50$\,\% up to a temperature of twice the value of $T_{\rm N}$, and a significant negative \textit{MR} is observed at even higher temperatures in rather large fields (measured up to 11\,T). As mentioned above, in fields of order 1\,T, a significant negative \textit{MR} sets in at about 150\,K, i.e. six times T$_N$. Strikingly, this temperature roughly coincides with the temperature where the magnetic susceptibility at $\mu_0H = 1$\,T shows deviations from a Curie-Weiss law, see Fig.~\ref{Inverse_Sus_EuZn2P2.png}. In other materials, such a behavior has been understood as the onset temperature of the formation of ferromagnetic clusters (magnetic polarons) \cite{AleCrivillero2023},
which become stabilized in systems with low carrier density and strong magnetic exchange between local Eu$^{2+}$ moments and charge carrier spins. Magnetic polarons grow in size upon lowering the temperature and couple to external (or internal) magnetic fields such that the mobility of the localized carriers, which can be estimated to $\mu \approx 8\,{\rm cm^2/Vs}$ at room temperature and zero magnetic field, see Sec.~\ref{sec:infrared}, becomes drastically enhanced thereby explaining large negative \textit{MR}. If such a scenario involving  electronic and magnetic phase separation as well as strong coupling to the lattice degrees of freedom \cite{Das2012, Manna2014}
 applies to EuZn$_2$P$_2$ will be the subject of future investigations.

	\subsection{Magnetic measurements}
    \vspace*{-\baselineskip}
 \begin{figure}[ht]
		\centering
		\includegraphics[width=0.47\textwidth]{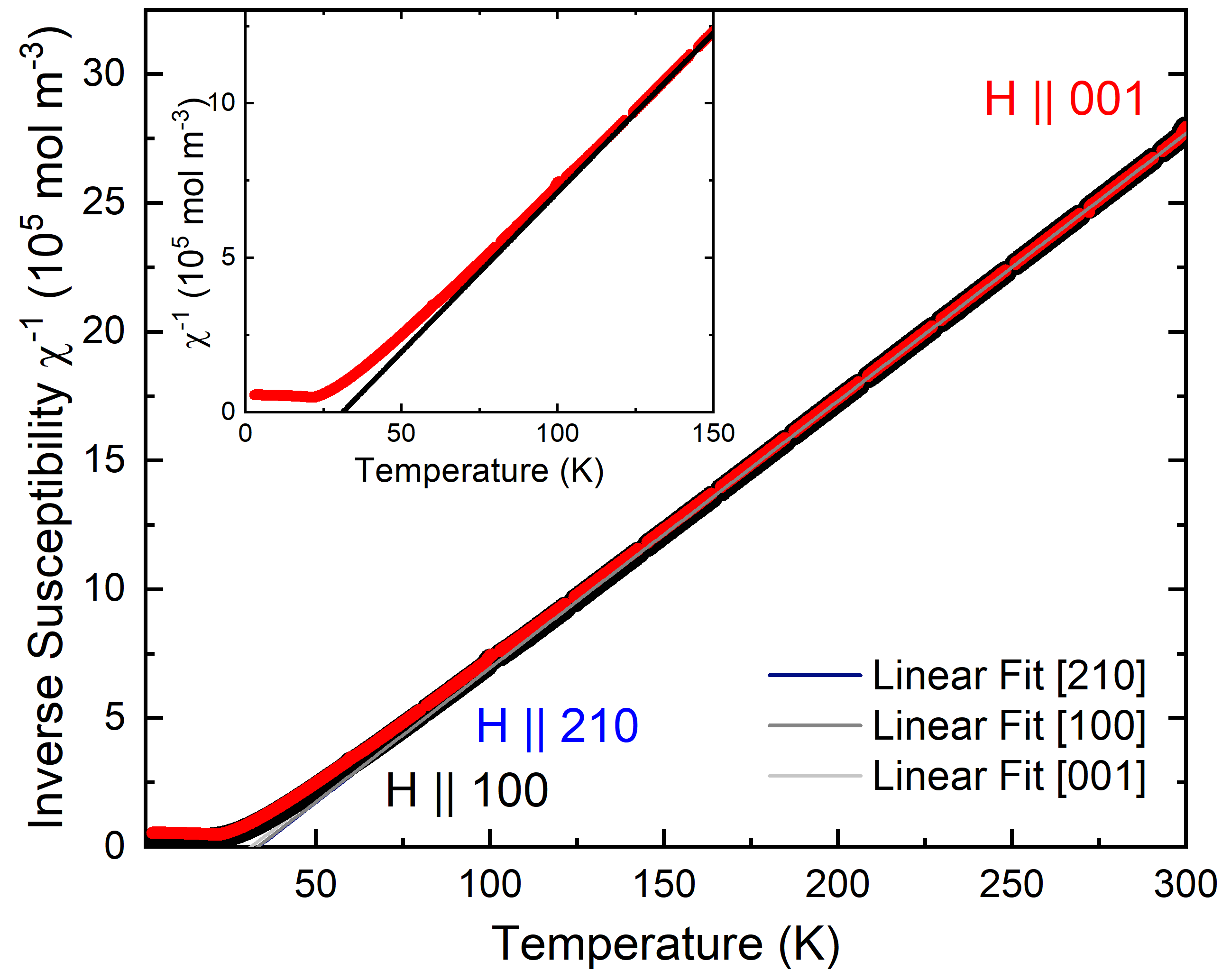}
		\caption[]{Inverse of the magnetic susceptibility as a function of temperature for all crystallographic orientations measured at $\mu_0H = 1\,\rm T$. The lines show the Curie–Weiss fit at high temperatures. Inset: Inverse magnetic susceptibility for $\mu_0H = 1\,\rm T$ for H || [001].}
		\label{Inverse_Sus_EuZn2P2.png}
	\end{figure}

 \begin{figure*}[ht]
			\includegraphics[width=0.42\textwidth, keepaspectratio]{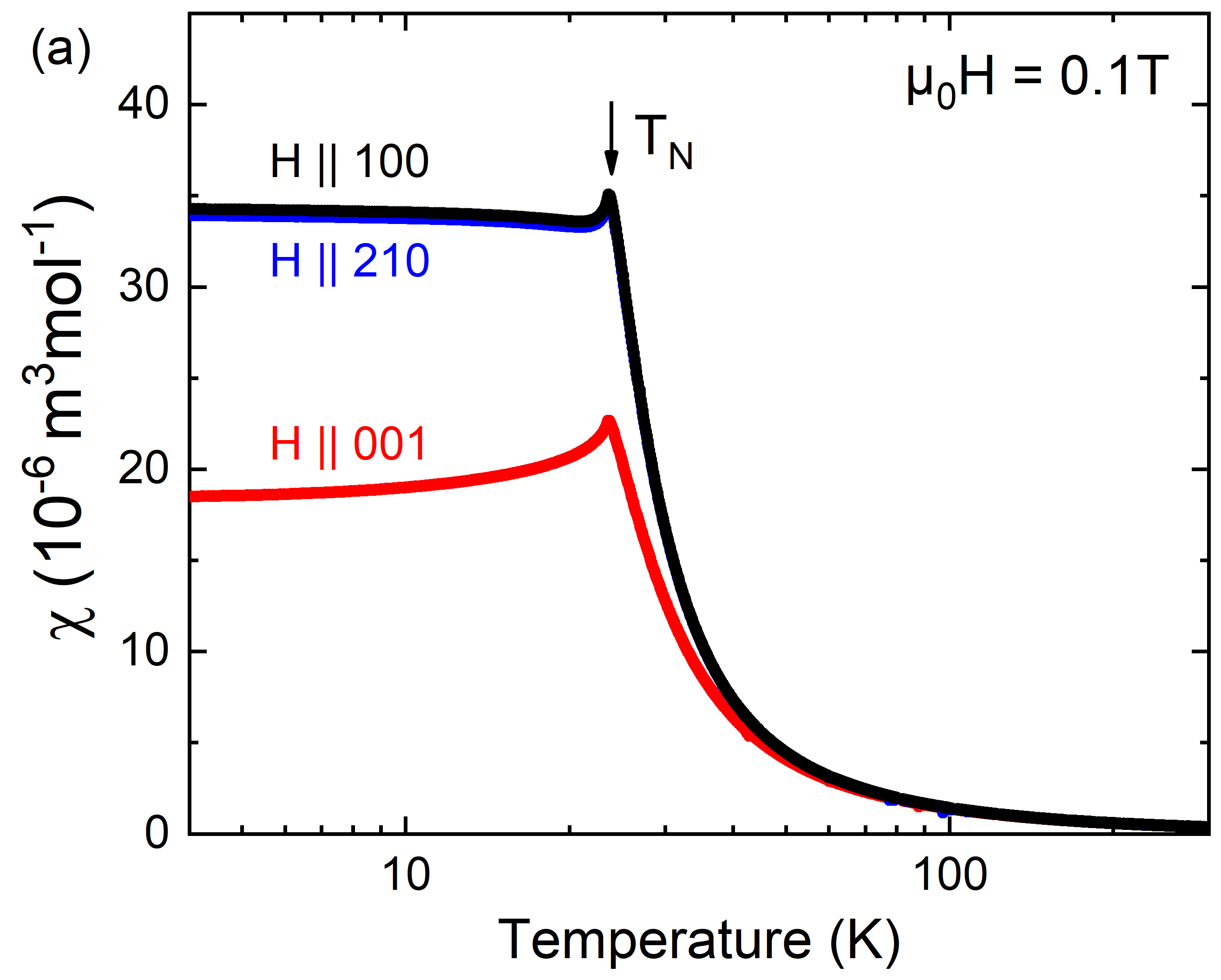}
			\includegraphics[width=0.42\textwidth, keepaspectratio]{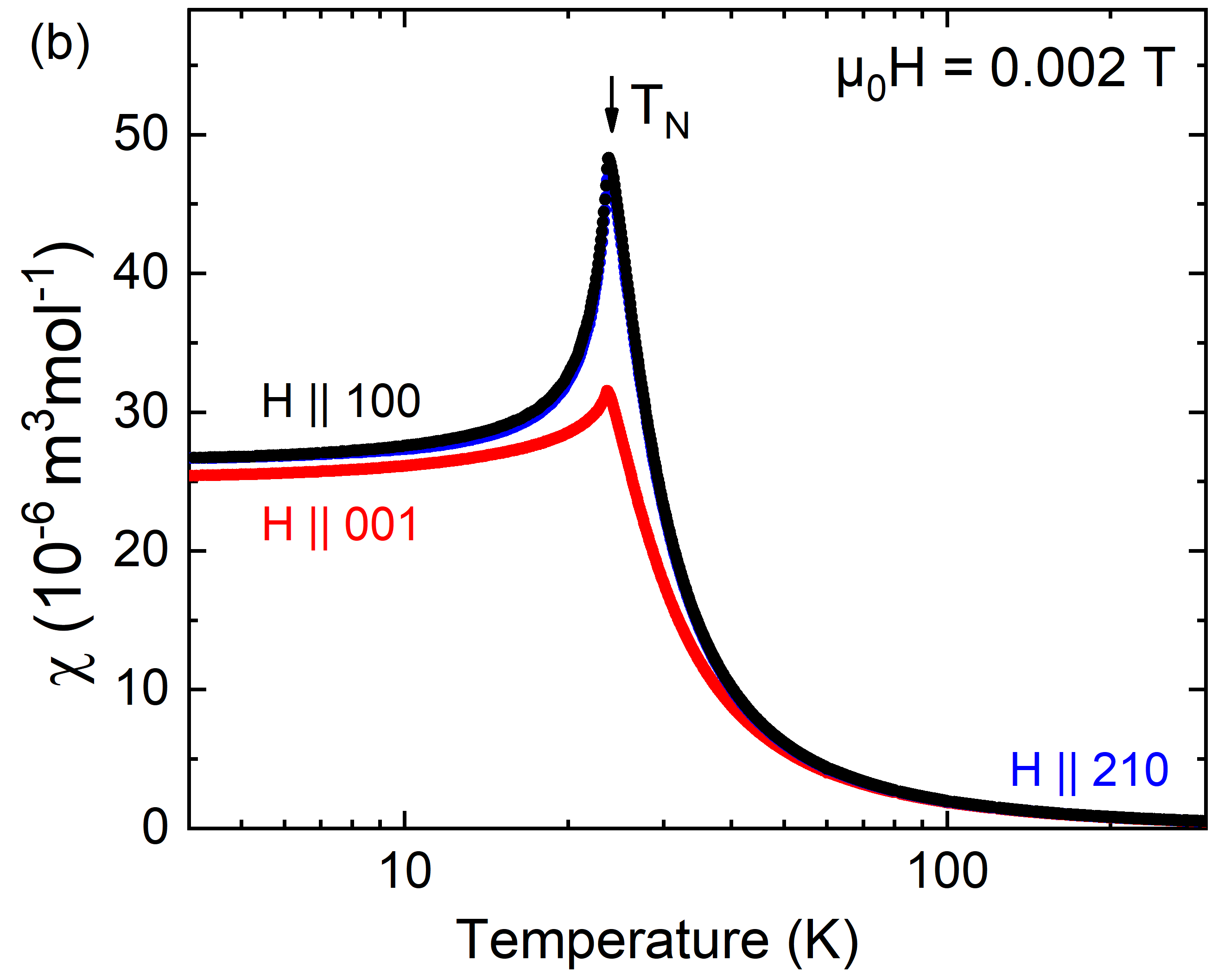}
			\includegraphics[width=0.42\textwidth]{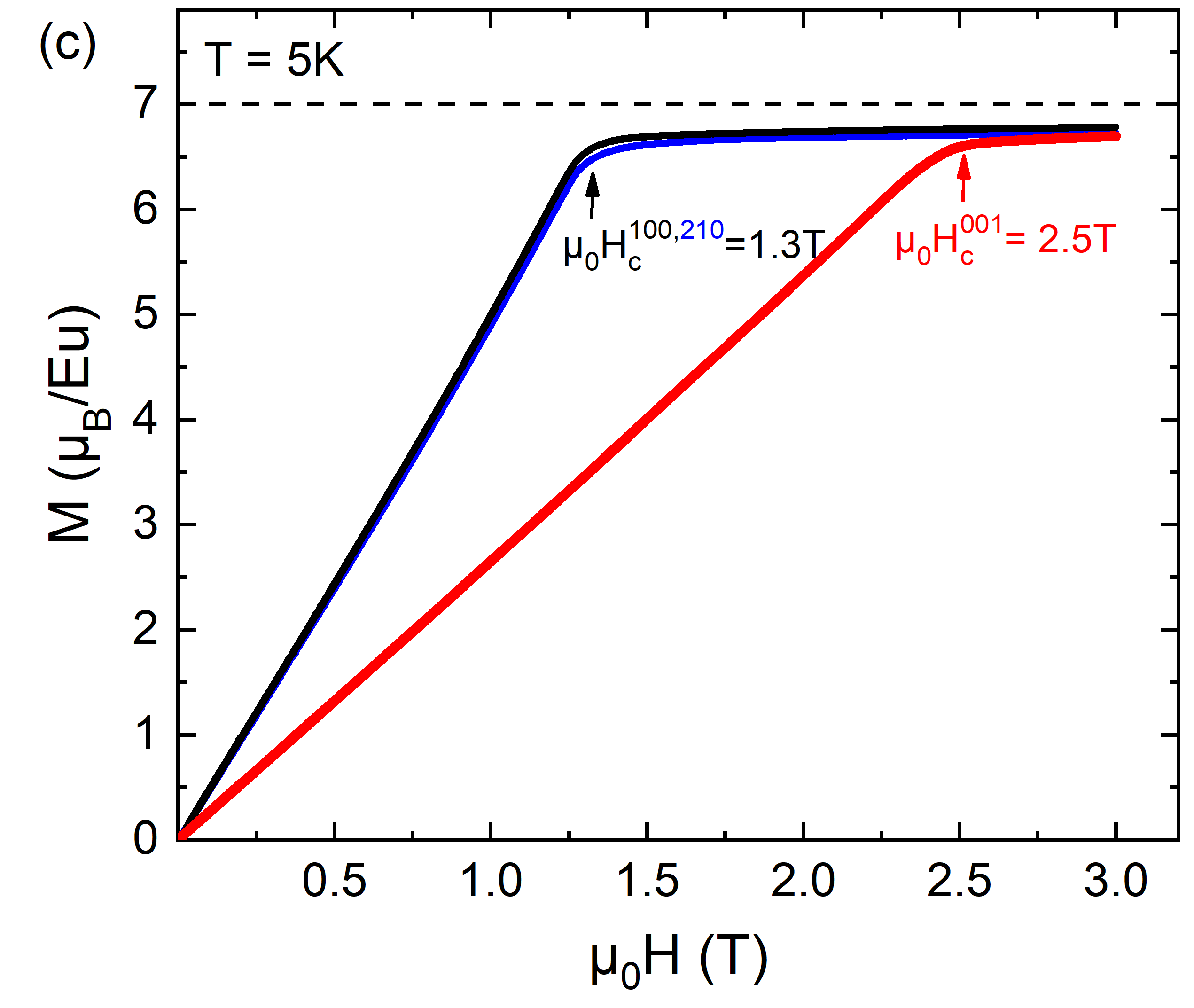}
			\includegraphics[width=0.42\textwidth]{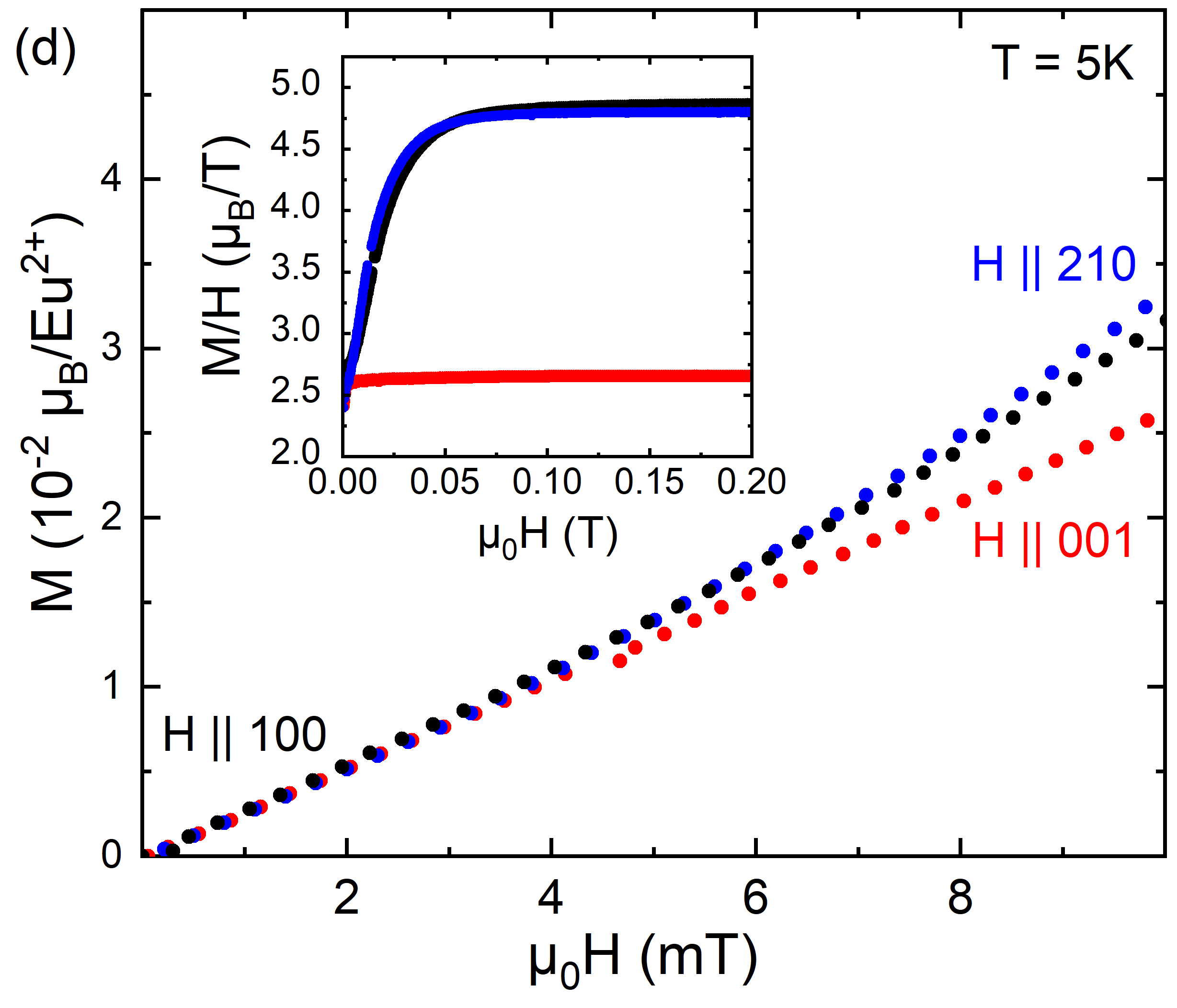}
		\caption[]{Magnetic susceptibility of EuZn$_2$P$_2$ at (a) $\mu_0H = 0.1\,\rm T$ and (b) $\mu_0H = 0.002\,\rm T$. (c) Field-dependent magnetization of EuZn$_2$P$_2$ for both in-plane directions and the out-of-plane direction and for (d) small magnetic fields. Inset: $M/\mu_0H$ as a function of $\mu_0H$.}
		\label{Isotropy_of_EuZn}
	\end{figure*}
	
The field-dependent magnetization $M(H)$ as well as the temperature-dependent magnetic susceptibility $\chi(T)$ was measured with an applied magnetic field along the $[001]$-, $[100]$- and $[210]$- directions, see Figs.~\ref{Isotropy_of_EuZn}(a-d). The obtained results, shown in the supplemental information in Figs.~\ref{MvT_MvH_001} and \ref{PD_001_100_210}, are similar to those reported in Ref.~\cite{Berry2022} but according to the results of our careful Laue analysis, we find that the critical field to reach the field-polarized state is smaller for in-plane directions, compared to the magnetic field along c, Fig.~\ref{Isotropy_of_EuZn}(c). In Ref.~\cite{Berry2022} the opposite case was proposed.
In order to fully represent the magnetic ground state, small magnetic fields, and low temperatures need to be considered. This is particularly important in Eu-based systems, where magnetic anisotropies due to spin-orbit coupling are rather weak and small magnetic fields can easily polarize the localized moments. Therefore, the temperature-dependent magnetic susceptibility was measured at $\mu_0H = 0.1\,\rm T$ shown in  Fig.~\ref{Isotropy_of_EuZn}(a) and at $\mu_0H = 0.002\,\rm T$, shown in Fig.~\ref{Isotropy_of_EuZn}(b). While the magnetic susceptibility is anisotropic for $\mu_0H = 0.1\,\rm T$, we find that at very low magnetic fields and at low temperatures the magnetic anisotropy becomes nearly indistinguishable. At $\mu_0H = 0.002\,\rm T$, the data for $H\parallel [001]$ (red) and $H\perp [001]$ are almost isotropic. Especially for field in-plane, $H\parallel [100]$ (black) and $H\parallel [210]$ (blue) an anisotropy is absent. For higher magnetic fields, we observe a pronounced anisotropy between the in-plane and the out-of-plane directions as presented in  Fig.~\ref{Isotropy_of_EuZn}(c) for $T=5\,\rm K$ with fields applied along the three crystallographic main symmetry directions. The magnetization saturates at approximately $7\,$\textmu$_{\rm B}$, which is the expected value for divalent Eu$^{2+}$. While for $H\parallel [100]$ (black) and $H\parallel [210]$ (blue) the saturation is reached already at $\mu_0H_c^{100,210}=1.3\,\rm T$, a higher $\mu_0H_c^{001}=2.5\,\rm T$ is observed for $H\parallel [001]$ (red). From the low-field magnetization, information about the orientation of the magnetic moments in a material can be obtained, as recently demonstrated in detail for a tetragonal system \cite{Kliemt2017}. In the case of trigonal  EuZn$_2$P$_2$, with a lower in-plane symmetry and where the angular difference between the $[100]$ and the $[210]$ directions is smaller than in the  tetragonal material (30 degrees instead of 45 degrees), we find two different types of magnetization curves for different samples as shown in the appendix in Fig.~\ref{domains_in_ChiandMvH}(b). Curves of type 1 show a close-to-linear behavior down to the lowest fields. Curves of type 2 show a large slope below $\mu_0H = 0.05\,\rm T$ which becomes smaller above this field value. As these changes occur at very low fields, we believe that this different magnetization behavior is due to the presence of AFM domains in the material. 
Fig.~\ref{Isotropy_of_EuZn}(d) emphasizes that the field-dependent magnetization is almost isotropic below magnetic fields of $\mu_0H = 5\,\rm mT$. This suggests a more complex magnetic structure compared to a simple $A$-type AFM structure with the magnetic moments lying in the $a-a$ plane. From $M/H$ versus $H$, shown in the inset in Fig.~\ref{Isotropy_of_EuZn}(d), we conclude that small fields applied along $H\parallel [100]$ or $H\parallel [210]$ are sufficient to induce a reorientation of AFM domains in this material. This low-field behavior together with the almost isotropic $\chi(T)$ at low temperatures hints at a moment orientation in the crystal structure on the one hand in between the $[100]$ and $[210]$ directions and on the other hand canted out-of-the plane towards the $[001]$ direction. This differs from the suggested alignment of the magnetic moments in the $a-a$ plane as shown in Fig.~5c of Ref.~\cite{Berry2022}. Using magnetization measurements, we cannot determine the exact orientation of the magnetic moments in the unit cell as we always average over all possible magnetic domains. Instead, we used resonant magnetic scattering where only one magnetic domain can be probed at the same time to determine the magnetic structure of the material. The results are shown in Sec.~\ref{sec:magscat}.
Fig.~\ref{Inverse_Sus_EuZn2P2.png} shows the inverse magnetic susceptibility of EuZn$_2$P$_2$ between 2\,K and 300\,K, where the data was fitted according to the Curie–Weiss law between 150\,K and 300\,K. The Weiss temperature $\Theta_W=(33\pm1)\,\rm K$ as well as the effective magnetic moment $\mu_{\rm eff}=(7.84\pm0.02)\,\mu_{\rm B}$ were determined, which are isotropic for the three crystallographic main symmetry directions. The effective moment agrees well with the theoretical value $\mu_{\rm eff}^{\rm theo}= 7.94\,\mu_{\rm B}$.
Interestingly, the inverse magnetic susceptibility shows a graduate deviation from the straight line of the fit starting already at high temperatures, however becoming significantly high below $\approx 150\,\rm K$. Due to the absence of orbital angular momentum ($L=0$) in Eu$^{2+}$, no crystalline electric field (CEF) effects and therefore no strong deviation from the Curie-Weiss fit is expected far above the magnetic ordering temperature. This deviation hints at the presence of magnetic fluctuations above $T_{\rm N}$. 

	\subsection{Magnetic structure determined from resonant magnetic X-ray diffraction}\label{sec:magscat}
    \vspace*{-\baselineskip}
	\begin{figure}
		\includegraphics[width=0.47\textwidth]{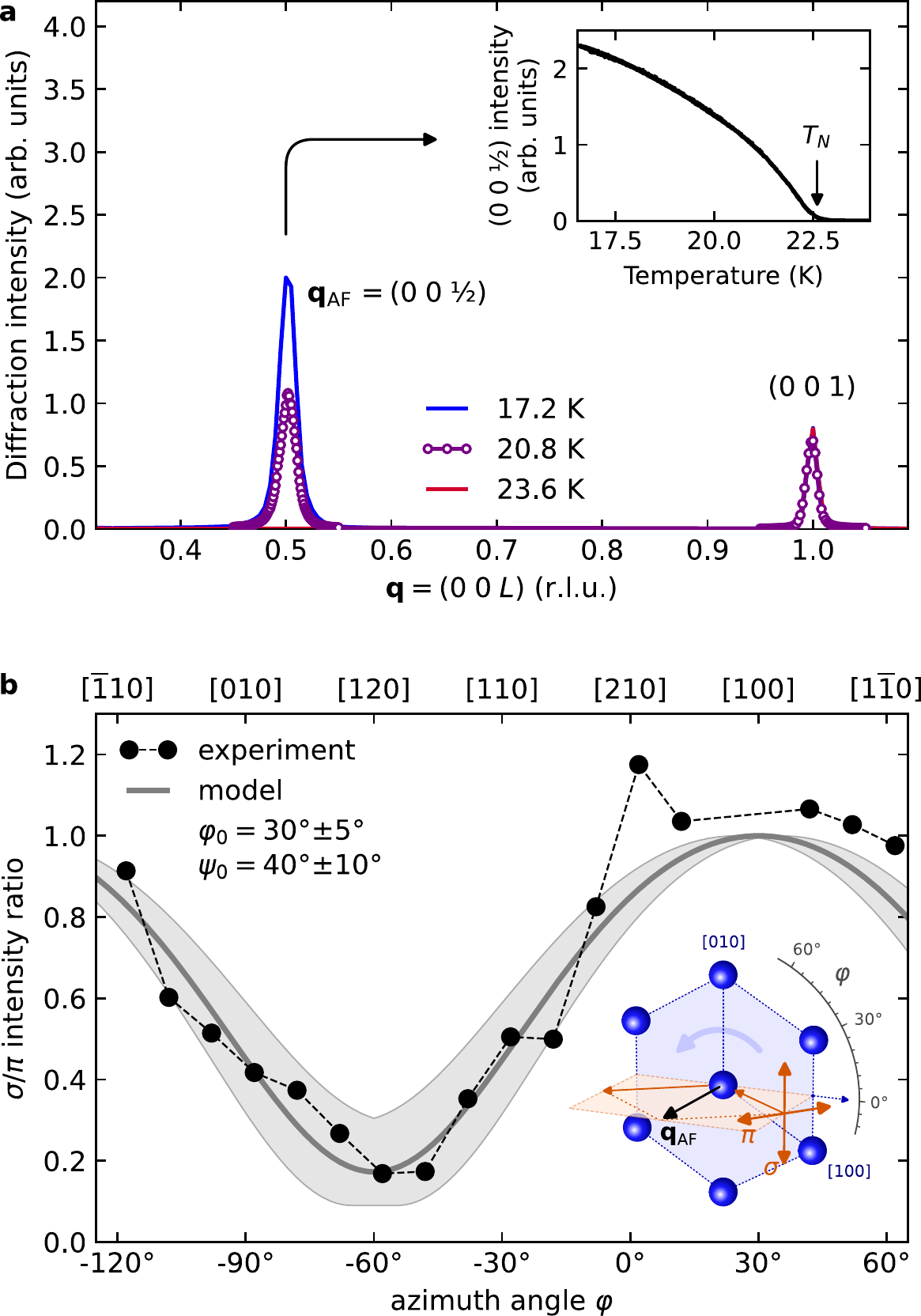}
		\caption{\label{fig:rmxd}(\textbf{a}) Magnetic Bragg peak at $\mathbf{q}_{\mathrm{AF}}=(0\;0\;\frac{1}{2})$ observed at the Eu $M_5$ edge for $T<T_{\rm N}$. The inset shows the evolution of the intensity of the $(0\;0\;\frac{1}{2})$ peak as $T$ is scanned across $T_{\rm N}$ confirming the magnetic origin of this peak. (\textbf{b}) Intensity ratio of the $(0\;0\;\frac{1}{2})$ peak for incident $\sigma$ vs $\pi$ polarisation as the sample is rotated around the $(0\;0\;\frac{1}{2})$ direction. The bottom axis shows the rotation angle while the top axis indicates which crystallographic direction is rotated into the scattering plane for this azimuth angle. The best fit to the observed azimuth dependence within a simple A-type AFM model is obtained for an orientation of the magnetic moments along the Eu-Eu direction in the $a-a$ plane and a canting by $40^{\circ}\pm 10^{\circ}$ away from the $a-a$ plane.}
	\end{figure}
 
	In order to determine the magnetic structure of EuZn\textsubscript{2}P\textsubscript{2} we performed resonant magnetic diffraction \cite{hill1996-ac} at the Eu $M_5$ edge. Tuning the incident energy to the Eu $3d\rightarrow 4f$ $M_5$ resonance during the diffraction experiment enhances the magnetic diffraction peaks originating from the AFM order of the Eu moments by several orders of magnitude such that they become comparable in intensity to lattice Bragg peaks. The intensity of magnetic Bragg peaks strongly depends on the relative orientation of the magnetic moments with respect to the incident light polarisation and the diffraction plane. Here, we took advantage of these characteristics to study the ordering of the magnetic Eu moments.
	Fig.~\ref{fig:rmxd}(a) shows a series of scans along $(00L)$ performed below and above $T_{\rm N}$. In addition to the usual $(001)$ structural Bragg peak, we observe the emergence of a magnetic Bragg peak at $(0\;0\;\frac{1}{2})$ when the temperature is lowered below $T_{\rm N}$. The presence of a magnetic peak at $\mathbf{q_{AF}}=(0\;0\;\frac{1}{2})$ is indicative of a magnetic unit cell twice the size along $c$ compared to the structural unit cell, as is the case for a simple '+\,-\,+\,-' antiferromagnetic stacking of ferromagnetically ordered Eu planes ($A$-type AFM order). Fig.~\ref{fig:rmxd}(b) shows the intensity ratio of the magnetic Bragg peak for incident $\sigma$ and $\pi$ polarisation as the sample is rotated about the [0 0 1] direction by an azimuth angle $\varphi$. The experimental geometry is shown as an inset. If the magnetic moments are aligned along the $c$-axis a rotation about $\varphi$ does not change anything in the relative orientation of the magnetic moments and the polarisation vectors of the incident light. In this case, the $\sigma/\pi$ intensity ratio will not be dependent on $\varphi$ and take a constant value of one everywhere. By contrast, for moments lying in the $a-a$ plane a strong modulation of the $\sigma/\pi$ ratio will occur between a value of one when the magnetic moments are in the diffraction plane and zero when they are perpendicular to the diffraction plane. For moments canted away from the $a-a$ plane a rotation about the $c$-axis cannot perfectly align the moments perpendicular to the diffraction plane anymore and the $\sigma/\pi$ ratio will reach its minimum at some finite value that depends on the canting angle $\psi_0$, away from the $a-a$ plane towards c. We performed a fit of a collinear $A$-type AFM model to the experimental data. From that, we found that the best agreement is achieved for the magnetic moments aligned along the Eu-Eu direction within the $a-a$ plane and a canting by about $40^{\circ}\pm 10^{\circ}$ out of the $a-a$ plane in agreement with an in-plane component as suggested in Ref. \cite{Berry2022}.
	  The observation that the magnetic Bragg peak remains unchanged upon cooling, Fig.~\ref{fig:rmxd}(a), indicates the absence of lattice distortion in EuZn$_2$P$_2$, unlike the Jahn-Teller distortion, e.g., in LaMnO$_3$, which is suggested to be related to the CMR effect in mixed-valence perovskite manganites \cite{Solovyev1996, Dagotto2001}. 
 
\subsection{DFT+U and ARPES} \label{sec:dft}\label{sec:arpes_dft}
\begin{figure*}[ht]
		\begin{center} 
				\centering
				\includegraphics[width=1.0\textwidth]{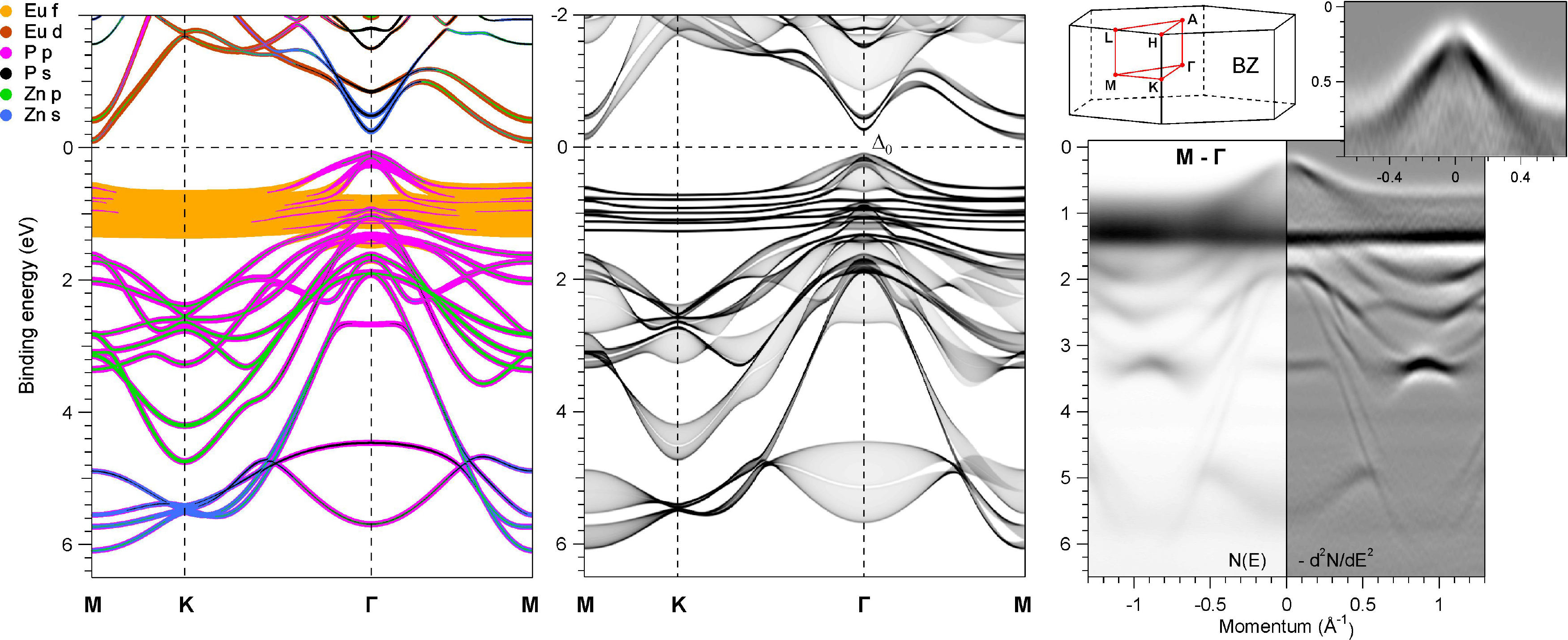}
				\caption[]{DFT+U and ARPES insight into the electronic structure of EuZn$_2$P$_2$. (a) Calculated orbital-resolved bulk electronic structure in the antiferromagnetic out-of-plane spin configuration along M - K - $\Gamma$ - M direction of the Brillouin zone (BZ). (b) The surface-projected bulk electronic structure on the (001) surface along the $\overline{\mathrm{M}}$ - $\overline{\mathrm{K}}$ - $\overline{\Gamma}$ - $\overline{\mathrm{M}}$ direction shown as grey-scale density of states. $\Delta_0$ denotes the direct band gap at the $\Gamma$ point.} (c) Experimentally derived electronic structure obtained using the photons of 137~eV along the $\overline{\mathrm{M}}$ - $\overline{\Gamma}$ - $\overline{\mathrm{M}}$ direction. The insets in (c) show the sketch of the Brillouin zone as well as the details of the spectral pattern of a cone-like feature at the $\overline{\Gamma}$ point.
				\label{EuZn_DFT}
		\end{center}
	\end{figure*}

    \vspace*{-\baselineskip}
    For the electronic characterization, we performed calculations in the framework of density functional theory (DFT) employing the Hubbard $U$ correction for the Eu~4$f$ states. Investigating both in-plane and out-of-plane AFM order of Eu 4$f$ moments revealed no considerable changes in the electronic band structure with only tiny changes observed in the structure of the Eu 4$f$ states. Because of this weak anisotropy, we will focus below on the results obtained for the out-of-plane orientation of the 4$f$ moments. The corresponding calculated band structure and its orbital composition are shown in Fig.~\ref{EuZn_DFT}(a) for $U=1.8$\,eV. Two prominent and characteristic features are visible, the first relating to the band gap, and the second to the flat Eu~$4f$ band. Starting with the band gap around the Fermi level $E_\mathrm{F}$, we can see that an indirect band gap  of 0.2\,eV is formed between the valence band maximum at the $\Gamma$ point and the conduction band minimum at the M point. The gap is slightly smaller than the direct band gap at $\Gamma$ of 0.34\,eV, thus the calculation predicts EuZn$_2$P$_2$ to be an indirect semiconductor. 
    Looking at the orbital composition, the excitation of the direct band gap is a transition from a cone-shaped state at the $\Gamma$ point, reminiscent of Dirac matter, dominated by P~$p$ states to an electron-like band dominated by Zn~$s$ and P~$s$ states, while the indirect excitation is a transition to primarily unoccupied Eu~5$d$ states. Hence, the energy of the Eu~5$d$ states in comparison to the Eu~4$f$ plays an important role for the size of the indirect band gap. Additionally, the Eu~4$f$ states which can be clearly seen as flat bands located at $\approx 1.0\,\rm eV$ below $E_\mathrm{F}$ (orange), hybridize with the cone-shaped bands that form the valence band maximum at the $\Gamma$ point. The Eu~4$f$ states intermix with the P~$p$-dominated states leading to an $f-p$ hybrid which, however, loses rapidly 4$f$ character away from the Eu 4$f$ states. Still, a small magnetic moment at the P sites ($\sim0.05$\textmu$_\mathrm{B}$) is induced. Similarly, $f-p$ hybrids have been observed in the AFM semiconductor EuCd$_2$As$_2$~\cite{Soh2019, Ma2019, Ma2020} or ferromagnetic semiconductor EuS~\cite{Fedorov2021}, where the hybridization strength and degree of Eu~4$f$ correlation were found to be crucial for the magnetic exchange interaction and size of the band gap.	
    To examine the validity of our DFT+U calculation, we performed ARPES measurements of the [001] surface of EuZn$_2$P$_2$ shown in Fig.~\ref{EuZn_DFT}(c), which can be best compared to the projection of the calculated bulk bands along $k_z$ as depicted in Fig.~\ref{EuZn_DFT}(b). In accordance with the electronic transport measurements and DFT+U calculation, the photoemission measurements show clear semiconducting properties with a visible gap at $E_\mathrm{F}$ and electrical charging of the sample at low temperatures. Using a photon energy of 137\,eV, both the valence band as well as the Eu~4$f$ states are clearly visible, which nicely agrees with the projected band structure. Especially, the agreement of the energy position of the Eu 4$f$ states in relation to the cone-shaped valence band maximum gives a good indication that the empirically chosen Hubbard $U=1.8$\,eV correction should allow for a good prediction of the unoccupied states, too. Further prominently visible in the ARPES data is the aforementioned $f-p$ hybridization of the Eu~4$f$ level with the P~$p$-dominated cone-like feature around the $\Gamma$ point. The spectral structure of the cone-like feature is shown as an inset in  Fig.~\ref{EuZn_DFT}(c) with the valence band maximum having a typical Dirac-like shape. Finally, the survey ARPES spectrum (not shown) taken from a freshly cleaved EuZn$_2$P$_2$ single crystal does not reveal any traces of the Eu$^{3+}$ final state multiplet situated between 6 and $12\,\rm eV$. This is in clear agreement with the purely divalent magnetic configuration of Eu expected for EuZn$_2$P$_2$.
	
\subsection{Infrared spectroscopy}\label{sec:infrared}
\begin{figure}
	\centering
	\includegraphics[width=0.5\textwidth]{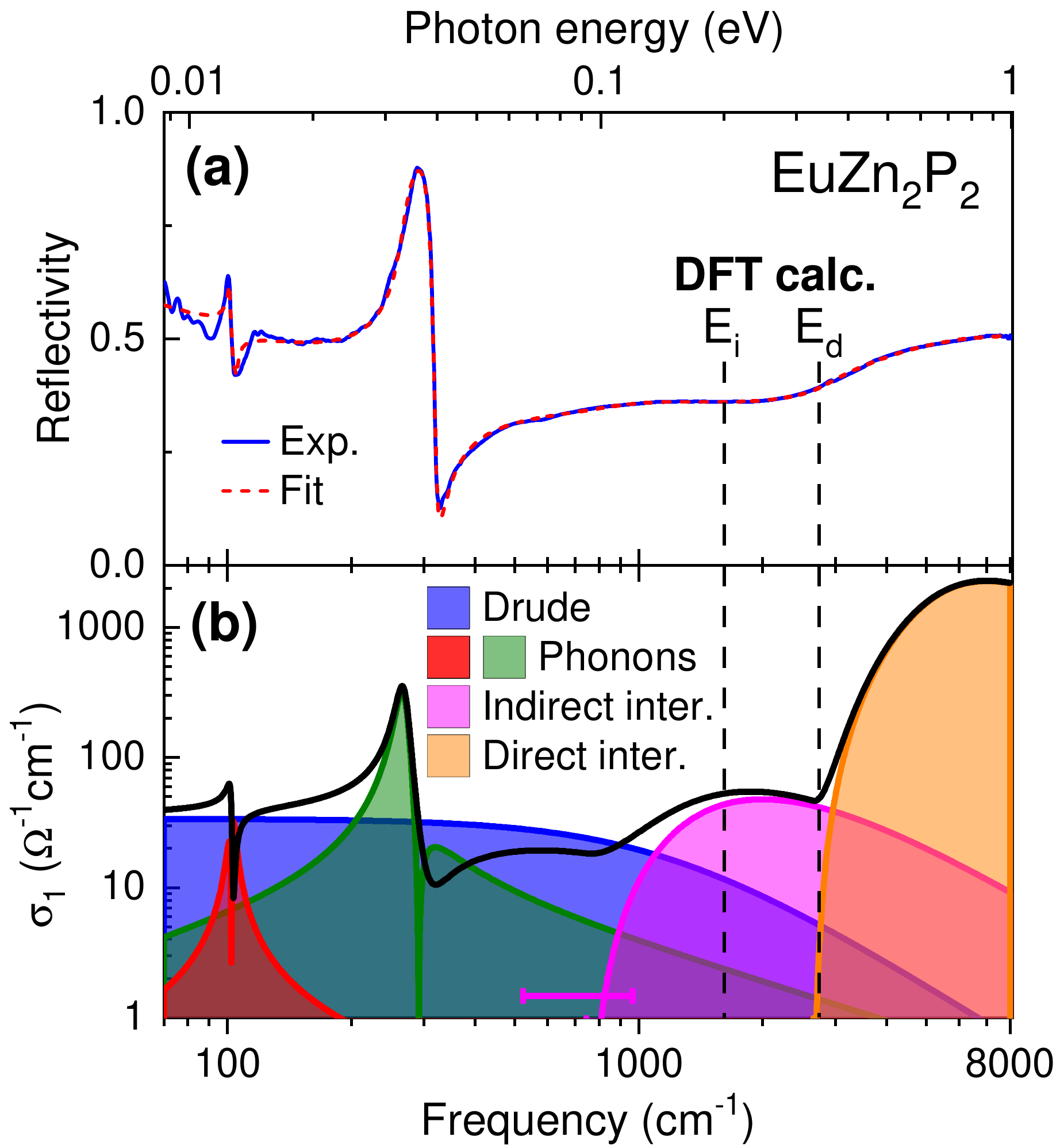}
	\caption{(a) Near-normal incidence reflectivity spectrum $R(\omega)$ of \EuZnP at $T=295$\,K: experimental (blue curve) and model fit (red dashed curve, see text for details). (b) Model optical conductivity spectrum $\sigma_1(\omega)$ (black curve) used for fitting $R(\omega)$ in (a).  Contributions from Drude, phonon, and (in)direct interband transitions as indicated in the legend (see text for key band parameters). Also included for comparison are the direct ($E_d$) and indirect ($E_i$) band gap energies from DFT+U calculations Sec.~\ref{sec:dft}) as vertical lines.  The horizontal magenta bar indicates the error margin for $E_i$ from the fitting procedure.  Note that for the two Fano phonon band contributions, $|\sigma_1(\omega)|$ is plotted, with $\sigma_1(\omega)<0$ for the higher energy sides ($1/q<0$). \label{fig:ir_gs}}
\end{figure}

    \vspace*{-\baselineskip}
    In order to study the low-energy intra-/interband and lattice excitations, in particular, for comparison with the transport and band structure properties investigated above, we measured the ground-state reflectivity spectrum $R(\omega)$ at $T=295$~K for light polarized in the $a-a$ plane over a broad spectral range from 10\,meV to 1\,eV. Fig.~\ref{fig:ir_gs}(a) shows the measured spectrum, where one observes two sharp (phonon) signatures at lower energy (energies given below) on top of a weakly metallic plateau, while in the range ${\gtrsim}1000\pcm$ a broad structure with slowly increasing $R(\sigma)$ and an inflection near $2000\pcm$ is present. Away from the main phonon peak, the reflectivity is well below unity, which is consistent with \EuZnP being a poor conductor with a small charge carrier concentration (see below). While certain general features are similar to those previously reported spectra for \EuCdP \cite{Homes23} and \EuCdAs \cite{Wang2016}, there are also differences that we address below after modeling the spectrum. For the free-carrier Drude response, we employ a standard Drude model $\sigma_D$ with a DC conductivity $\sigma_{0D}$ and scattering time $\tau_D=1/\Gamma_D$. To model the phonon bands, as is evident from the oscillatory form of $R(\omega)$ (especially for the lower band), we found one must go beyond basic Lorentzian lineshapes $\sigma_L(\omega)$ and employ the more general Fano lineshape $\sigma_F(\omega)$ \cite{sedlmeier12}, which accounts for electron-phonon coupling with the free carriers (Sec.~\ref{sec:ir_details}), as was also employed to model the higher-energy phonon in \EuCdAs \cite{Wang2016} (although not for \EuCdP \cite{Homes23}). %
    For the interband transitions, we use a Tauc-Lorentz band model $\sigma_g(\omega)$ ($g=i,d$), which differs from the commonly used Lorentzian model by including a low-energy cut-off to reflect the bandgap energies $E_{g}=\hbar \omega_{g}$ \cite{jellison96,jellison1996erratum}.
    This band model has been applied e.g. in amorphous materials \cite{jellison96} and indeed also conforms to the band-edge response for indirect transitions \cite{dresselbook}, although we tentatively employ it here for the direct interband transition as well, in order to avoid a long low-energy extension of the band, as would be the case for a standard broad Lorentzian band.
    The resultant fitted reflectivity spectrum is shown in Fig.~\ref{fig:ir_gs}(a) (dashed curve) and is seen to be in very good agreement with experiment. The corresponding fitted conductivity spectrum is shown in Fig.~\ref{fig:ir_gs}(b), including each separate contribution.
    Inspecting first the lower-energy range, the Drude contribution has a fitted DC conductivity of $\sigma_{0D}=33.9\pOhmpcm$, comparable to but somewhat higher than from the room temperature DC transport measurements (where $\sigma_{DC}=1/\rho_{DC}=20.8\pOhmpcm$, Sec.~\ref{sec:transport}).
    Interestingly, a similar analysis of \EuCdP also had a discrepancy between these two values, although there is a larger factor ${\sim}2$ and in the opposite direction \cite{Homes23}, which may indicate the presence of additional contributions to the conductivity response at even lower energies than measured here. Note that we did not observe any clear indication for a non-Drude conductivity signature, as was found for \EuCdP below ${\sim}400\pcm$ \cite{Homes23}.
    The Drude response is seen to be very broad due to a rapid scattering rate $\tau_D=4.6$~fs, corresponding to a damping rate of $\Gamma_D=1163\pcm$, which is comparable to that found in \EuCdP ($\Gamma_D=700\pcm$ \cite{Homes23}), although for \EuCdAs this rate was significantly lower ($\Gamma_D=190\pcm$ \cite{Wang2016}). One can propose that these differences result from variations in the band structures of these isoelectronic compounds, which affect the acoustic-/optical-phonon momentum scattering rates \cite{lundstrombook}, although one cannot rule out the possible role of impurities. In any case, the Drude plasma frequencies $\omega_p=\sqrt{\sigma_{D0}\Gamma_D/\varepsilon_0}$ are all closer, here $\omega_p=1540\pcm$ compared to $900\pcm$ (\EuCdAs, \cite{Wang2016}) and $1200\pcm$ (\EuCdP, \cite{Homes23}), i.e. the total spectral weight and carrier-density:mass ratios ($N/m^{*}$) are comparable \cite{dresselbook}.

    Using the conductivity effective mass (deduced from the DFT band structure - see Sec.~\ref{sec:density}) we can estimate a carrier concentration of $N = m^{aa}\sigma_{0D}/e^2\tau_D = 2.8\cdot 10^{18}\pcmc$, i.e. a value somewhat higher than might be expected from the relatively low magnitude of $\sigma_{0D}$ alone (compared to typical semiconductors), the latter due mainly to the short scattering time $\tau_D$. We can compare this to the expected carrier density based on thermal excitation of intrinsic carriers across the indirect band gap (Sec.~\ref{sec:density}), which yields $N=1.5\cdot 10^{17}\pcmc$, i.e. more than an order of magnitude smaller.  This indeed suggests that additional donor states exist in the material with a smaller activation energy than the predicted band gap (see Sec.~\ref{sec:transport}).

    Turning to the two phonons, the fitted bands in Fig.~\ref{fig:ir_gs}(b) are at frequencies $\omega_{p1}=102\pcm$ and $\omega_{p2}=269\pcm$, close to those in \EuCdAs ($86\pcm,167\pcm$ \cite{Wang2016}) and \EuCdP ($89\pcm,239\pcm$ \cite{Homes23}), which have been assigned to $E_u$ modes for light polarized in the $a-a$ plane. Interestingly, while Fano effects due to electron-phonon coupling could be disregarded in \EuCdP \cite{Homes23}, and were only significant in \EuCdAs at lower $T$ \cite{Wang2016}, our fitted spectra indicate rather strong Fano effects, with $1/q_1=-1.2$ and $1/q_2=-0.25$ -- such magnitudes for $1/q$ have been observed in other non-Eu systems, e.g. \cite{sedlmeier12}. The resultant distorted lineshapes can be seen in Fig.~\ref{fig:ir_gs}(b), and as mentioned above, were necessary to achieve the detailed fit of $R(\omega)$ in these spectral regions. This intriguing result of a stronger electron-phonon coupling in \EuZnP compared to \EuCdP and \EuCdAs should be pursued in future work including temperature dependence and application of Kramers-Kronig using an extended frequency range, in order to study the experimental conductivity spectra directly.
	
Finally, we inspect the fitted interband transitions, which was a key motivation for the ground-state infrared study, in order to compare the results with the measured and calculated band structures (Sec.~\ref{sec:arpes_dft}). The indirect/direct interband conductivity bands ($\sigma_{i,d}$, respectively) are included in Fig.~\ref{fig:ir_gs}(b), along with the corresponding band-gap energies from the DFT+U calculations with $U=1.8\eVolt$ (vertical dashed lines). One sees that the fitted band gap for the direct interband transition ($E_d=0.33\eVolt$, $2650\pcm$) coincides very closely with the theoretical prediction of $E_d=0.34\eVolt$.
While such close agreement is most likely coincidental, given the inherent issues of estimating optical excitation energies from DFT band structures \cite{kirchner21}, the agreement provides support for the accuracy of the DFT+U calculations presented above.
Turning to the indirect band, while the relative discrepancy for $E_i$ is somewhat larger (optical: $E_i=0.09\eVolt$, DFT+U: $E_i=0.20\eVolt$), this still amounts to a rather small absolute energy difference. Due to the spectral overlap in the region about $1000\pcm$ and only weak features in the reflectivity $R(\omega)$, the $1\sigma$-uncertainty in the fitted $E_i$ value was found to be $\pm 0.03\eVolt$ (denoted by the horizontal bar in Fig.~\ref{fig:ir_gs}(b) about $E_i$), which may account for part of the discrepancy.  In any case, the relative spectral weights of the fitted bands are consistent with the predicted band structure, i.e. with a much weaker indirect band gap due to the lower cross-section for phonon-assisted optical transitions \cite{dresselbook}.
	
\section{Summary and Discussion}
\vspace*{-\baselineskip}
Single crystals of EuZn$_2$P$_2$ were grown from Sn flux and structurally, chemically and physically characterized. The trigonal crystal structure (space group No. 164) was confirmed with single crystal X-ray diffraction where the data is consistent with the previously reported in Ref.~\cite{Berry2022}. In Ref.~\cite{Berry2022}, insulating behaviour was seen when measured down to $140\,\rm K$. In contrast, we found that the electrical resistivity of EuZn$_2$P$_2$ shows a semiconducting behaviour at low temperatures with a flattening of the resistivity below $T_{\rm N}$. The resistivity can be strongly suppressed by applying magnetic fields up to $\Delta \rho/\rho$ of $-99.4$\,\% at $\mu_0H = 4$\,T and $-99.9$\,\% at $11$\,T (CMR-effect). The onset of a considerable negative \textit{MR} at about $150\,\rm K$ coincides with a deviation of the inverse susceptibility from a Curie-Weiss behaviour. This indicates that the formation of magnetic polarons might be at the origin of the here reported CMR in EuZn$_2$P$_2$ and may also play an important role for the compound EuCd$_2$P$_2$. 

Using resonant magnetic X-ray diffraction we studied a single antiferromagnetic domain of a EuZn$_2$P$_2$ crystal, determined a magnetic ordering vector $q = (0\,\, 0\,\, 1/2)$ and found that the moments are canted by $\psi_0=40^\circ\pm 10^\circ$ out of the $a-a$ plane and tilted towards the [100]-direction in the plane. This agrees with the magnetization data, where isotropic behaviour was observed experimentally for $H\rightarrow 0$ and $T\rightarrow 0$ for EuZn$_2$P$_2$. For small applied fields, below $0.05\,\rm T$, we observe a redistribution of magnetic domains which is similar to what was observed in EuCd$_2$As$_2$ \cite{Rahn2018}. At 0.1\,T, EuZn$_2$P$_2$ shows a distinct magnetic anisotropy of the in-plane and out-of-plane directions, which is consistent with the data shown by Ref \cite{Berry2022}. In agreement with the resistivity measurements, both DFT+U calculations and ARPES measurements show semiconducting properties in EuZn$_2$P$_2$. The good agreement between DFT+U band structure and ARPES pattern lets us theoretically predict an indirect gap of ${E_i^\mathrm{DFT}=0.2\,\rm eV}$ and a direct band gap of ${E_d^\mathrm{DFT}=0.34\,\rm eV}$, which corresponds very well with the experimentally determined band gaps $E_i^\mathrm{opt}=0.09\,\rm eV$ and $E_d^\mathrm{opt}=0.33\,\rm eV$ obtained through our infrared study. Our combined ARPES and DFT+U study further reveals a $f-p$ hybridization around the $\Gamma$ point which is common for magnetic Eu-based semiconductors and believed to be crucial for the exchange magnetic interaction and size of the band gap. Lastly, the comparison of the calculation for in- and out-of-plane antiferromagnetic Eu~4$f$ order reveals only tiny differences in the band structure without strong anisotropic behavior.

Finally, it is interesting to compare the physical properties of EuZn$_2$P$_2$ with other compounds of the same family where Zn is replaced by Cd and P is replaced by As or Sb. All compounds are isostructural and exhibit Eu antiferromagnetic order with similiar magnetic ordering temperatures ($T_{\rm N}$ = 19.2 K in EuZn$_2$As$_2$ \cite{Blawat2022}, $T_{\rm N}$ = 13.3 K in EuZn$_2$Sb$_2$ \cite{Weber2006}, $T_{\rm N}$ = 11 K in EuCd$_2$P$_2$ \cite{Schellenberg2011}, $T_{\rm N}$ = 9.5 K in EuCd$_2$As$_2$ \cite{Rahn2018}). The materials differ slightly in their unit cell volumes and the size of the spin-orbit coupling depends on the transition metal. Similar to the compounds EuCd$_2$P$_2$ \cite{Wang2021} and EuCd$_2$As$_2$ \cite{Gati2021}, EuZn$_2$P$_2$ displays a magnetic anisotropy of the in-plane and out-of-plane directions at magnetic fields above 0.01T. However, there is presently no evidence for a canting of the magnetic moments in EuCd$_2$P$_2$ away from the a-a plane \cite{Wang2021}. Furthermore, both phosphorous compounds show a strong colossal magnetoresistance at low temperatures \cite{Wang2021}. It would be interesting to investigate further whether the mechanism behind the CMR in EuZn$_2$P$_2$ and EuCd$_2$P$_2$ is in fact the same and whether the formation of magnetic polarons plays a role in the origin of the here reported CMR. 

Eu-based materials often show unusual magnetism, which is critical for nontrivial topological properties as seen in the paramagnetic phase of the compound EuCd$_2$As$_2$ \cite{Ma2019}. This compound is a candidate for a topological semimetal in which a Weyl phase is induced if Eu orders ferromagnetically with moments along the [001] direction \cite{Wang2019}. In the here investigated AFM sister compound EuZn$_2$P$_2$, we do not see any evidence of a topologically ordered state neither in the calculations nor in the ARPES data while studying the material above and below $T_{\rm N}$. However, we cannot fully exclude the possibility of its existence. While EuCd$_2$As$_2$ orders AFM with moments in the a-a plane, the magnetic moments are tilted out of plane towards the c direction in EuZn$_2$P$_2$. For EuCd$_2$As$_2$ it was shown that in-plane ferromagnetism can be induced by application of pressure \cite{Gati2021}. In analogy, it would be interesting to study the effect of pressure or strain on EuZn$_2$P$_2$ where the moments are canted out of plane already at ambient pressure.

\begin{acknowledgments}
\vspace*{-\baselineskip}
We thank T. F\"orster for technical support. We acknowledge funding by the Deutsche Forschungsgemeinschaft (DFG, German Research Foundation) via the TRR 288 (422213477, projects A03, A10, B02, B08). We acknowledge SOLARIS National Synchrotron Radiation Centre for the allocation of beamtime at the UARPES beamline. We acknowledge MAX IV Laboratory for time on the Bloch Beamline under Proposal 20211066. Research conducted at MAX IV, a Swedish national user facility, is supported by the Swedish Research council under contract 2018-07152, the Swedish Governmental Agency for Innovation Systems under contract 2018-04969, and Formas under contract 2019-02496.
\end{acknowledgments}

\bibliography{EuZn_Cd2P2}
\clearpage
\appendix
\noindent{\Large{\bf Appendix}}
	\setcounter{equation}{0}
	\renewcommand{\thesection}{S\arabic{section}}
	\renewcommand\thefigure{S\arabic{figure}}
	\setcounter{figure}{0} 

\section{Heat capacity}
\begin{figure*}[p]
	\begin{center} 
		\includegraphics[width=0.6\textwidth]{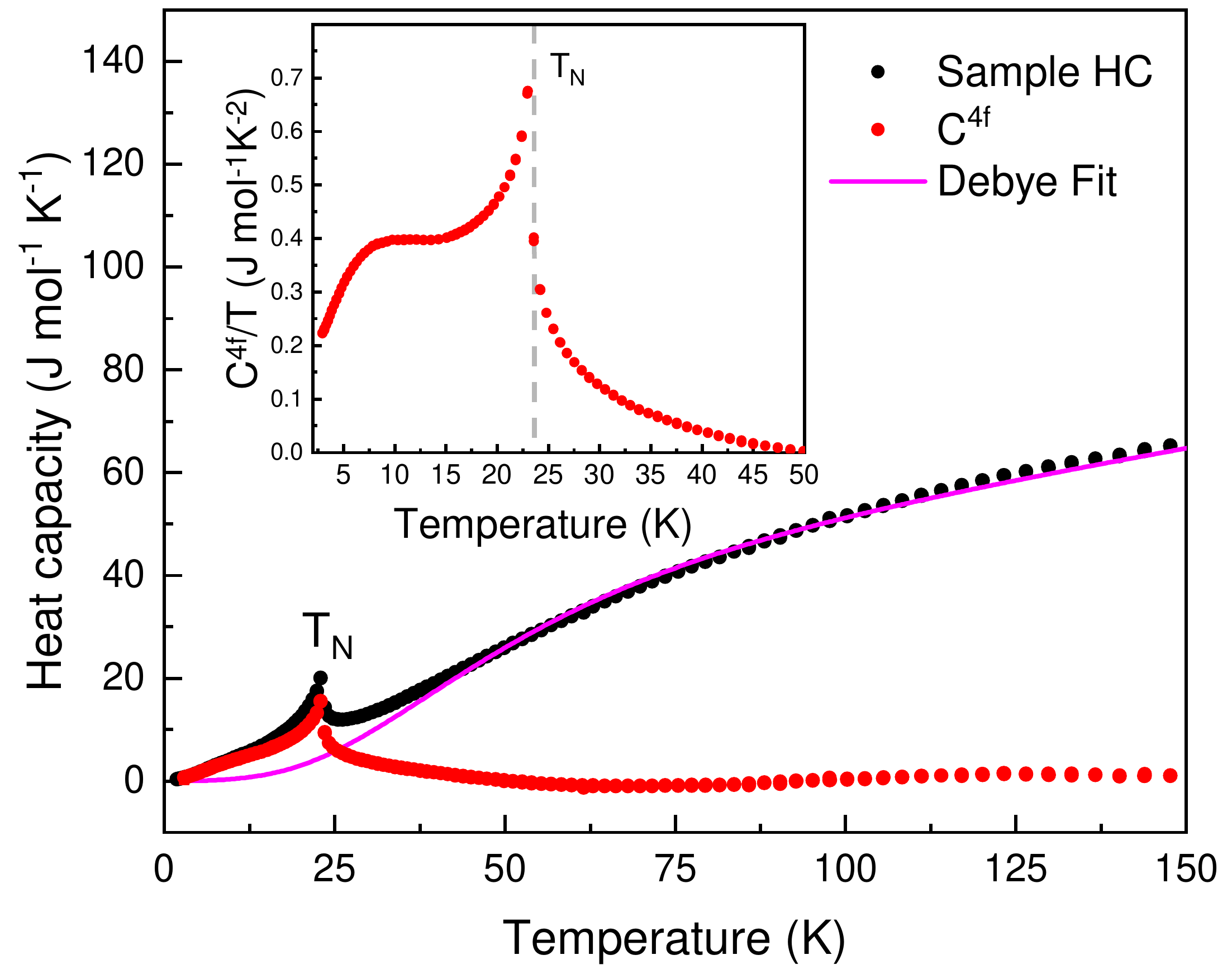}
	\end{center}
	\caption{Heat capacity of EuZn$_2$P$_2$ with $\mu_0H = 0\,\rm T$ (black). Using a Two-Debye model, the data was fitted (pink) in order to subtract the phonon contribution and to obtain the magnetic part $C^{4f}$ (red) of the heat capacity. Inset: Specific heat $C^{4f}$ divided by temperature.}
		\label{HC_EuZn2P2}
\end{figure*}

 \vspace*{-\baselineskip}
The heat capacity of EuZn$_2$P$_2$ was measured between $1.8\,\rm K - 200\,\rm K$, Fig.~\ref{HC_EuZn2P2}. It shows a sharp $\lambda$-type anomaly at the antiferromagnetic phase transition and is consistent with the data presented in \cite{Berry2022}. In order to subtract the phonon contribution at high temperatures, we fitted the data with a two-temperature Debye model with $\Theta_1 = 233.4\,\rm K$ and $\Theta_2 = 1164.1\,\rm K$ \cite{Anderson2019}. 
	
	\section{Electrical transport}
\begin{figure*}[p]
\centering
\includegraphics[width=0.6\textwidth]{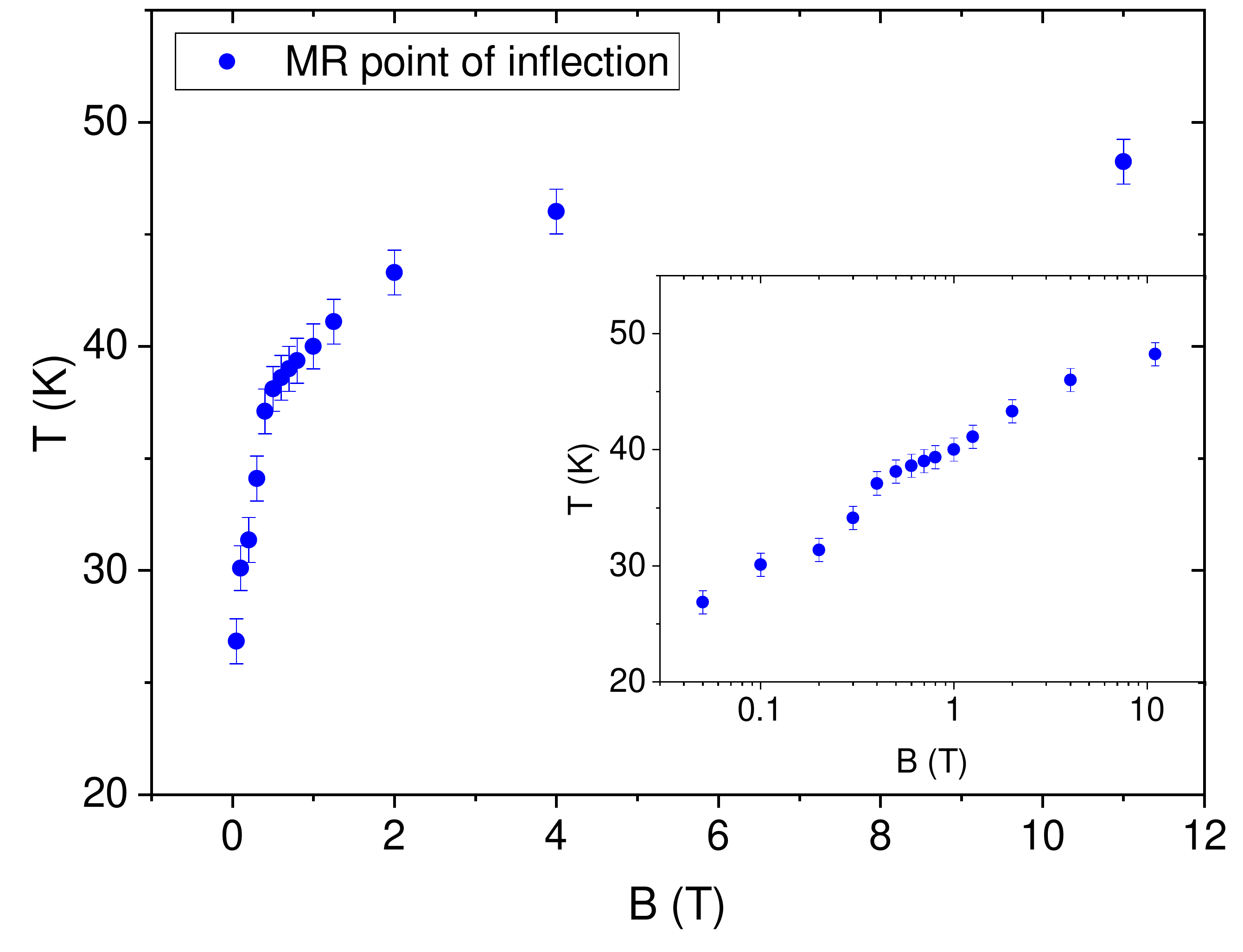}
\caption{Inflection point of the magnetoresistance curves versus $B=\mu_0H$. Inset: The same data plotted on a logarithmic scale. } 
\label{SI_inflection-point}%
\end{figure*}

  \vspace*{-\baselineskip}
The inflection point of the \textit{MR} vs. T curves marking the crossover to large suppression of the resistivity by magnetic fields is shown in Fig.~\ref{SI_inflection-point}. For fields $\mu_0H \gtrsim 0.5$\,T the \textit{MR} upon approaching $T_{\rm N}$ amount to more than $-90$\,\%. For rather large fields $\mu_0H \gtrsim 10$\,T the crossover temperature tends to saturate at about two times $T_{\rm N}$.

\section{Temperature dependence of the susceptibility}
\begin{figure*}[p]
\includegraphics[width=0.4\textwidth]{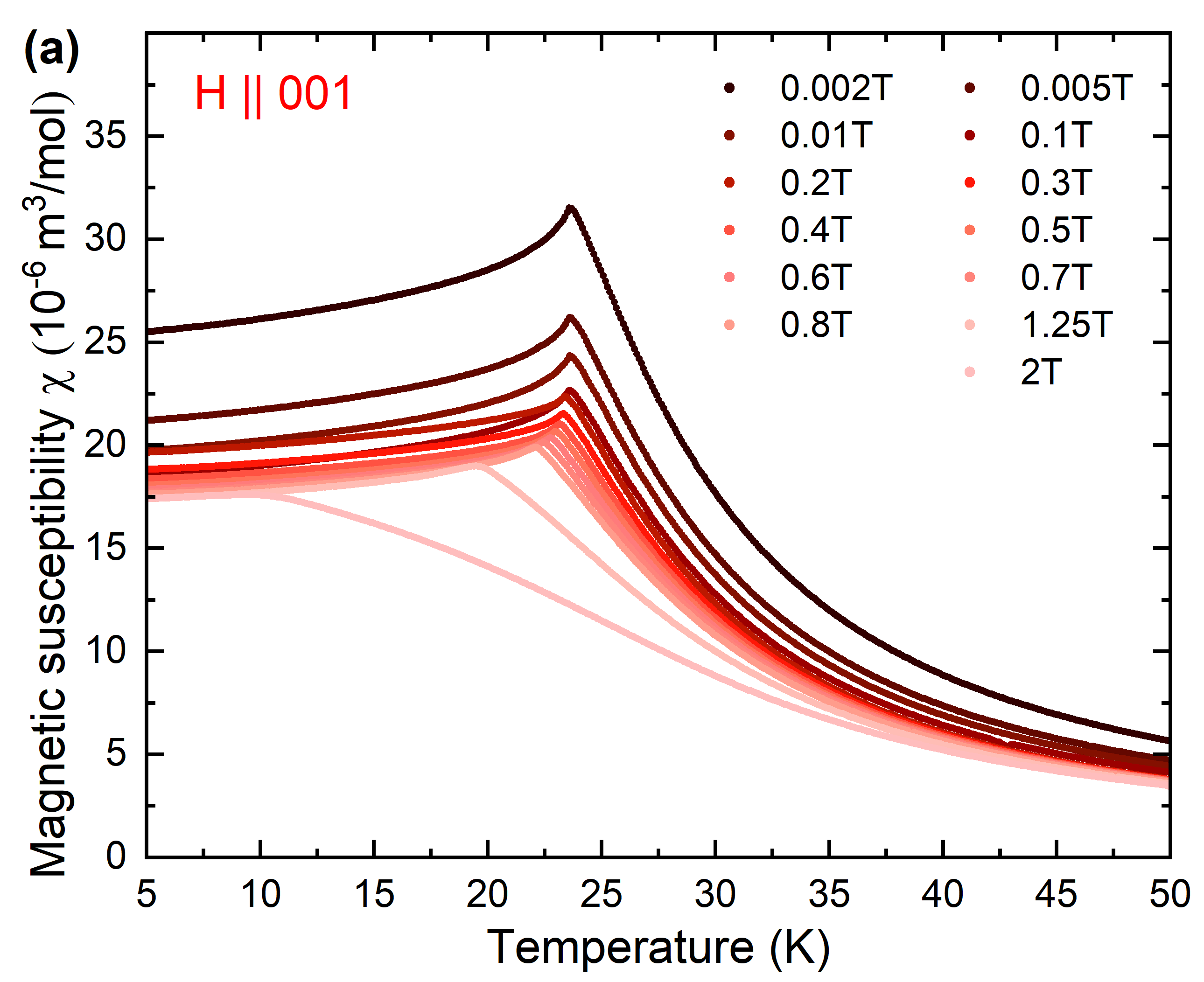}
\includegraphics[width=0.4\textwidth]{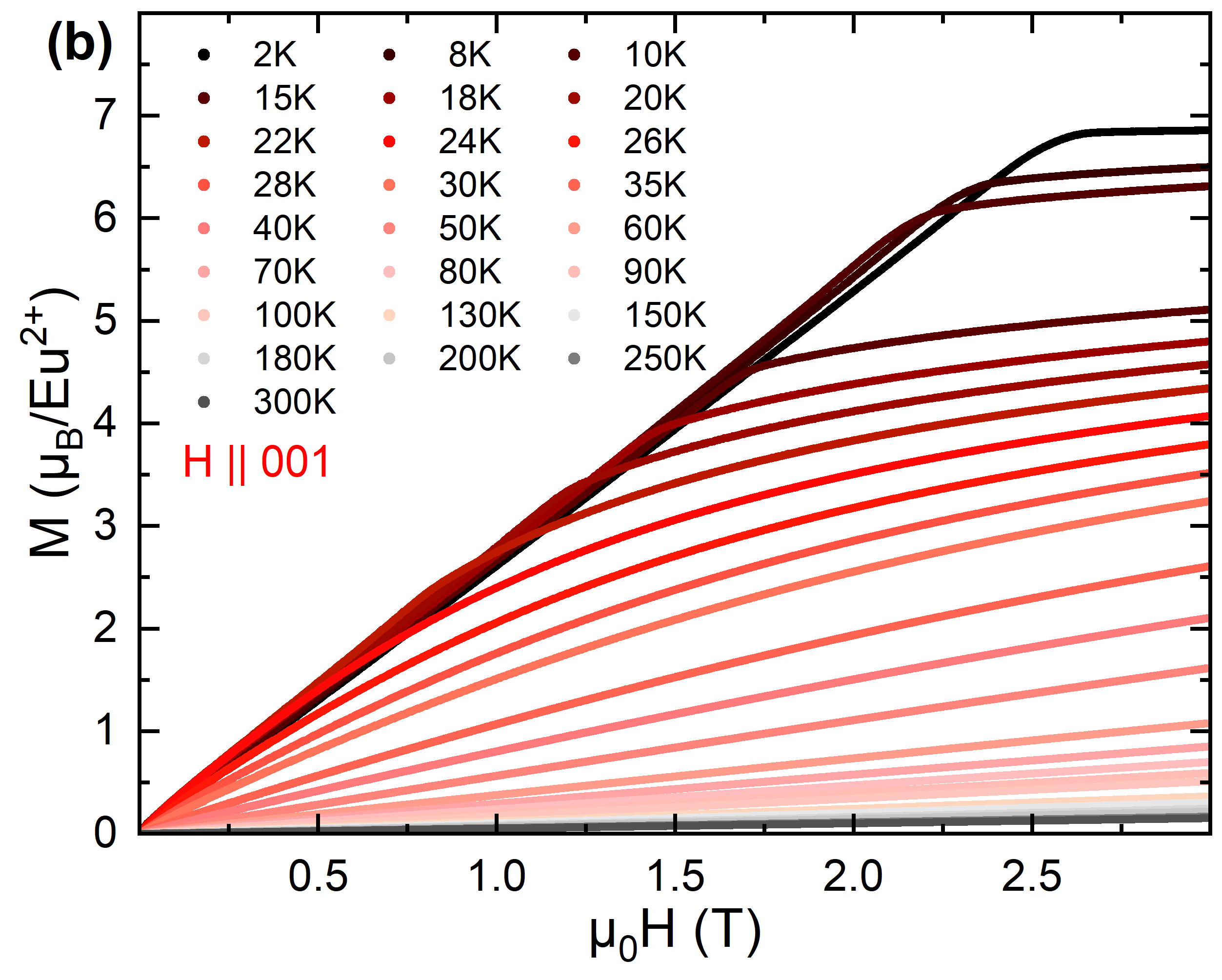}
\includegraphics[width=0.4\textwidth]{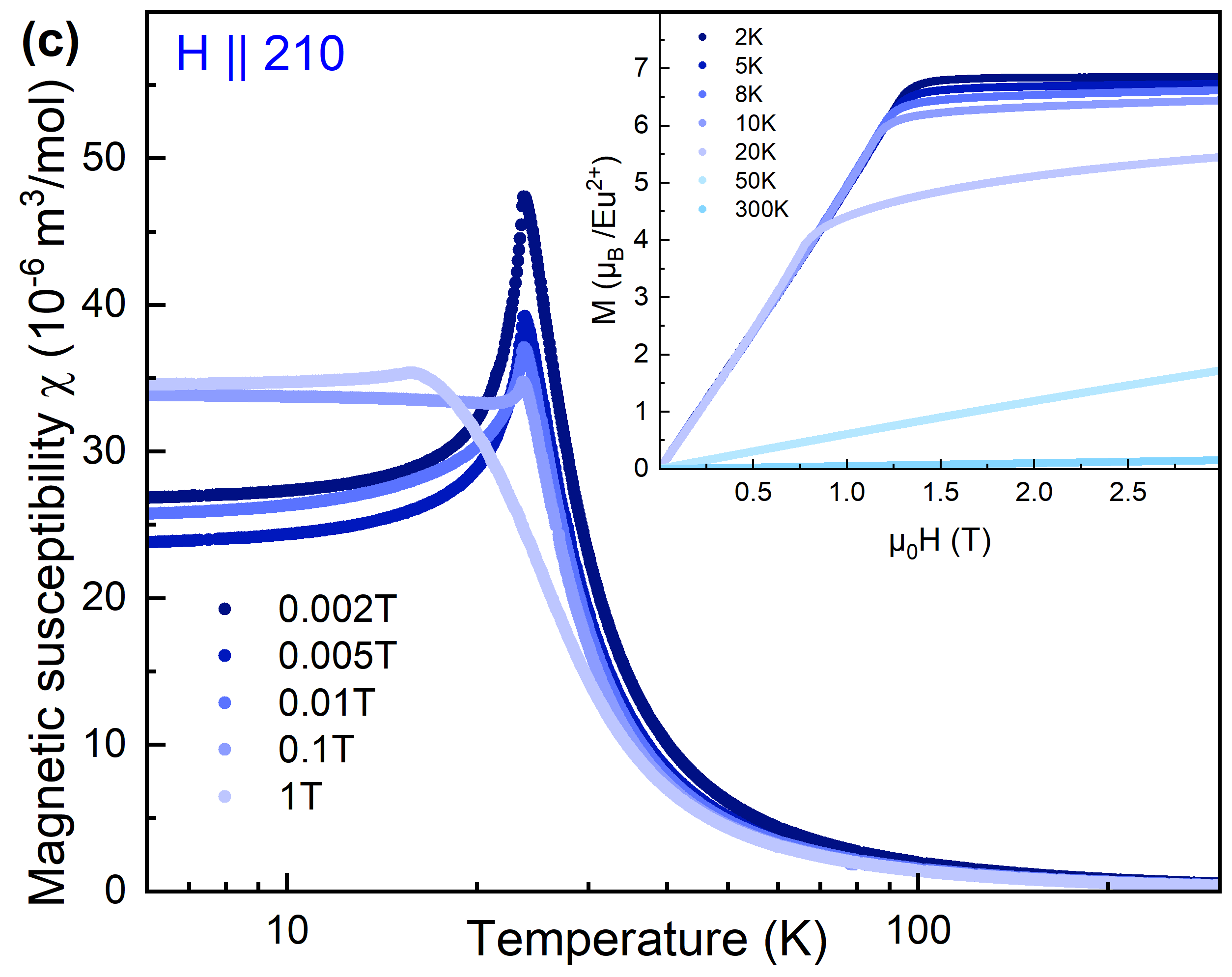}
\includegraphics[width=0.4\textwidth]{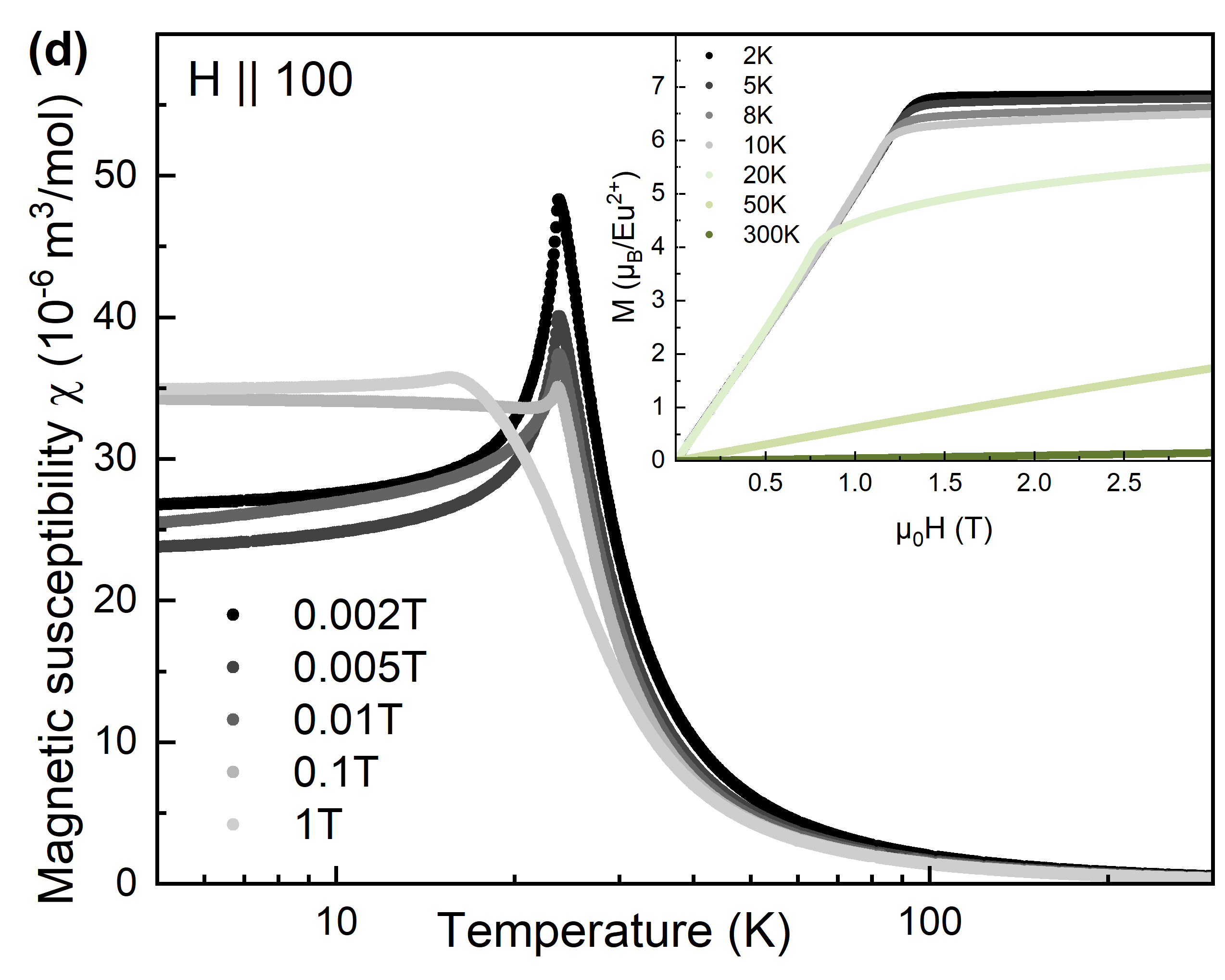}

\caption{Magnetic susceptibility as a function of temperature (a) and the field-dependent magnetization (b) for $H\parallel 001$. Magnetic susceptibility as a function of temperature for (c) $H\parallel 210$ and (d) $H\parallel 100$ for different magnetic fields. Insets: Field-dependent magnetization at different temperatures for [210] and [100] directions.}
	\label{MvT_MvH_001}
\end{figure*}

 \vspace*{-\baselineskip}
The bulk magnetic susceptibility of EuZn$_2$P$_2$ measured at different magnetic fields is shown in Fig.~\ref{MvT_MvH_001}. For all three main symmetry directions the antiferromagnetic phase transition occurs at $T_{\rm N}=23.7\,\rm K$ for low magnetic fields. For high magnetic fields, the transition temperature shifts towards lower values. 
	
\section{Field dependence of the magnetization}
\begin{figure*}[p]
		\begin{center} 
					\includegraphics[width=0.4\textwidth]{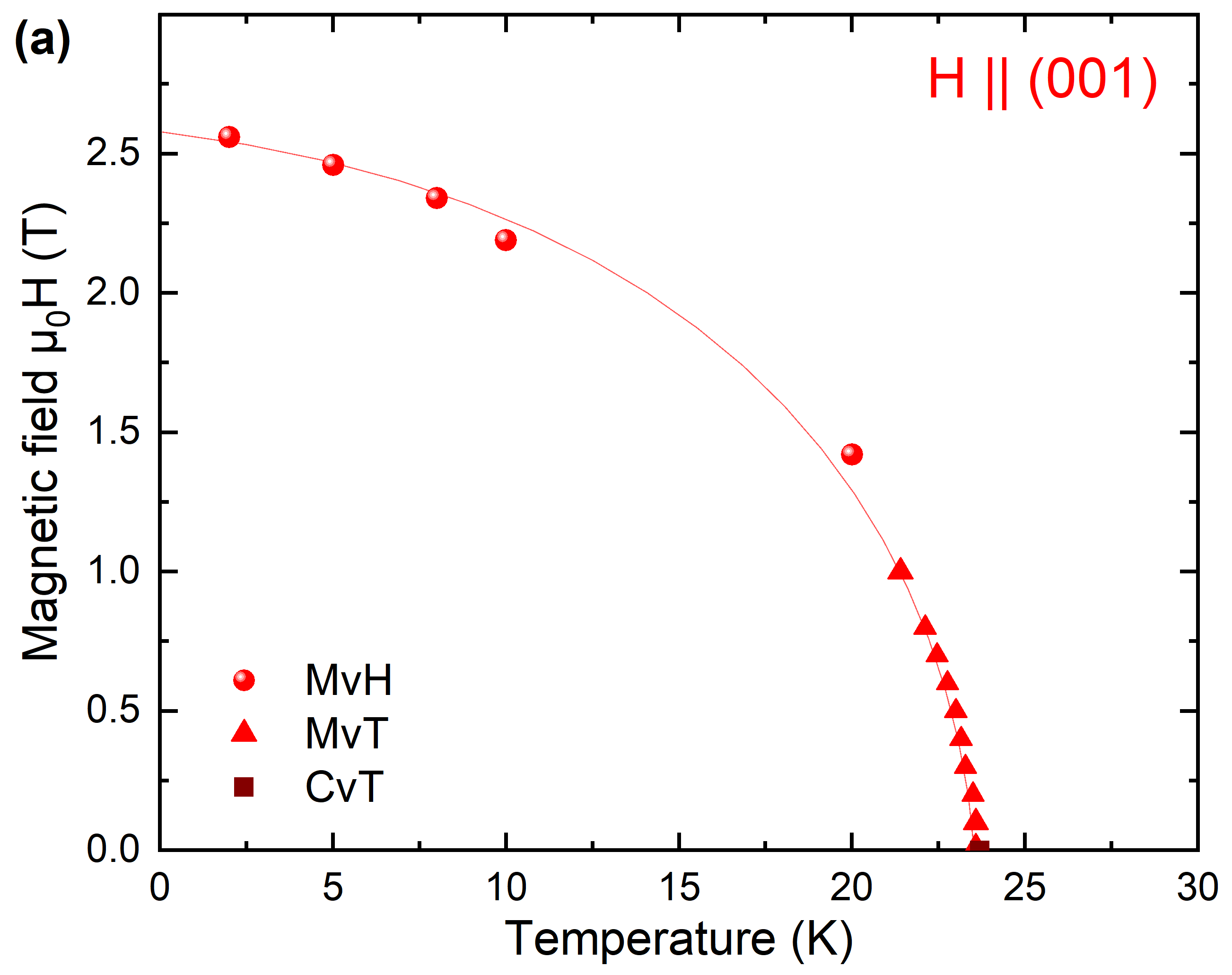}
					\includegraphics[width=0.4\textwidth]{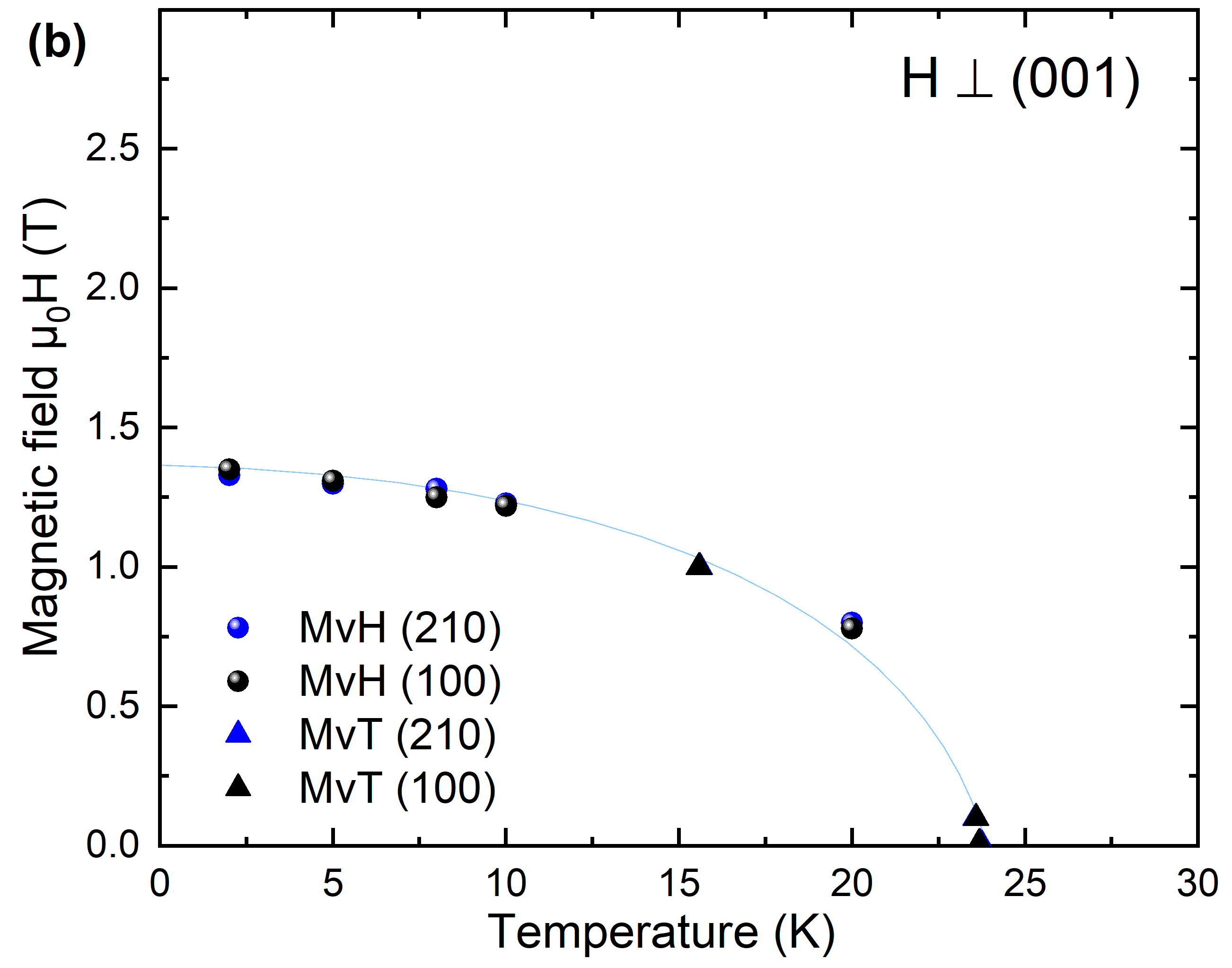}
		\end{center}
		\caption{Magnetic phase diagram of EuZn$_2$P$_2$ (a) for  the out-of-plane direction and (b) both in-plane directions. The data points were extracted from the magnetic susceptibility (MvT), the field-dependent magnetization (MvH), and the heat capacity (CvT).}
		\label{PD_001_100_210}
	\end{figure*}

 \vspace*{-\baselineskip}
The field-dependent magnetization measured at different temperatures shows that the magnetization saturates at $\approx 7\,\mu_B$ per Eu ion for all applied field directions, which is consistent with the divalent Eu$^{2+}$ configuration. However, the critical fields for saturation differ depending on the crystallographic direction. At $T=2\,\rm K$ for the $[001]$ direction, a larger field of $B_{c} = 2.57\,\rm T$ has to be applied to fully polarize the magnetic moments, see Fig.~\ref{MvT_MvH_001}(b). For the in-plane directions a magnetic field of only $B_{c} = 1.3\,\rm T$ is necessary, see Figs.~\ref{MvT_MvH_001}(c,d). The respective magnetic phase diagrams for all directions are shown in Fig.~\ref{PD_001_100_210}.

\section{EuZn$_2$P$_2$: Domain effects}
\begin{figure*}[p]
					\includegraphics[width=0.4\textwidth]{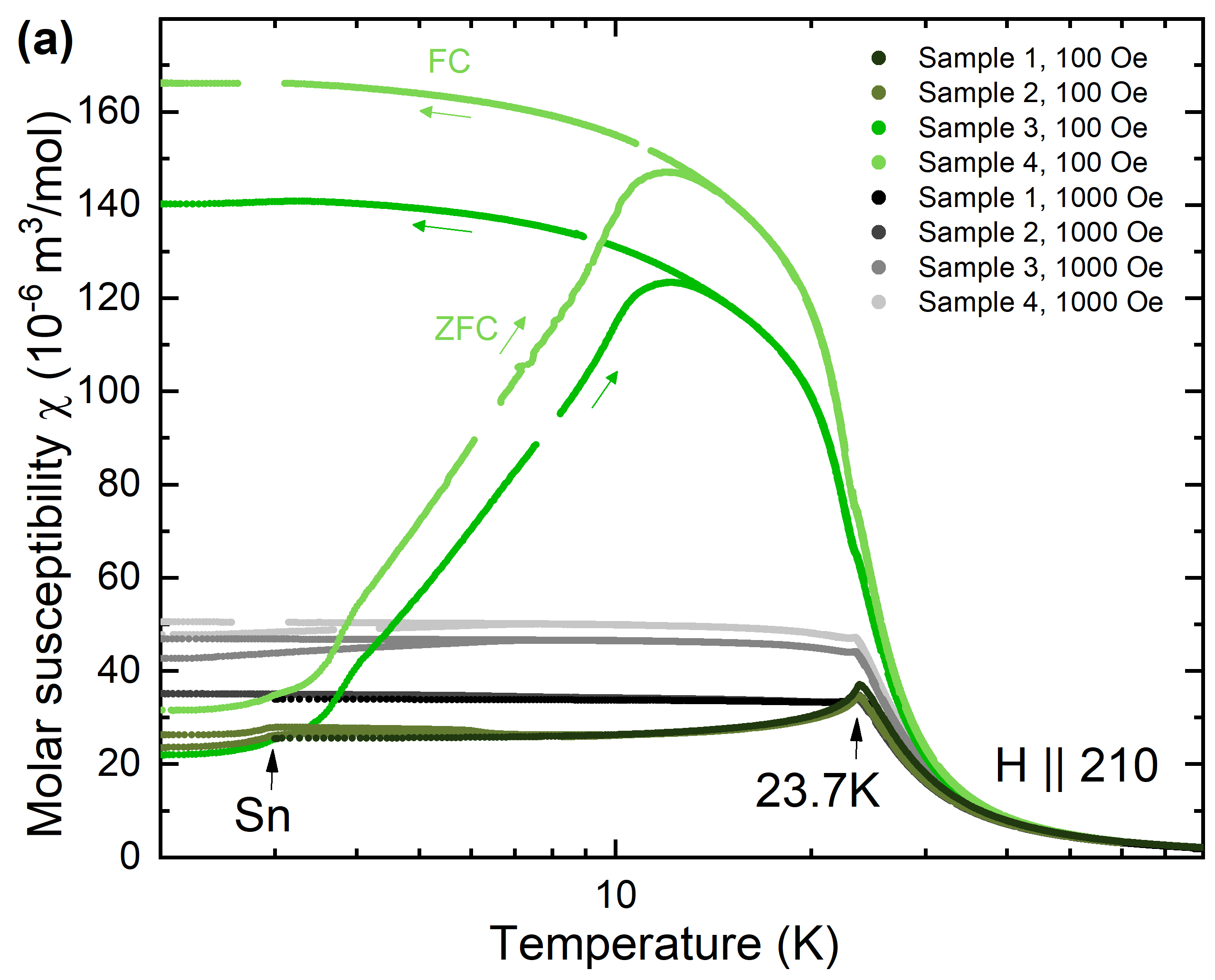}
					\includegraphics[width=0.4\textwidth]{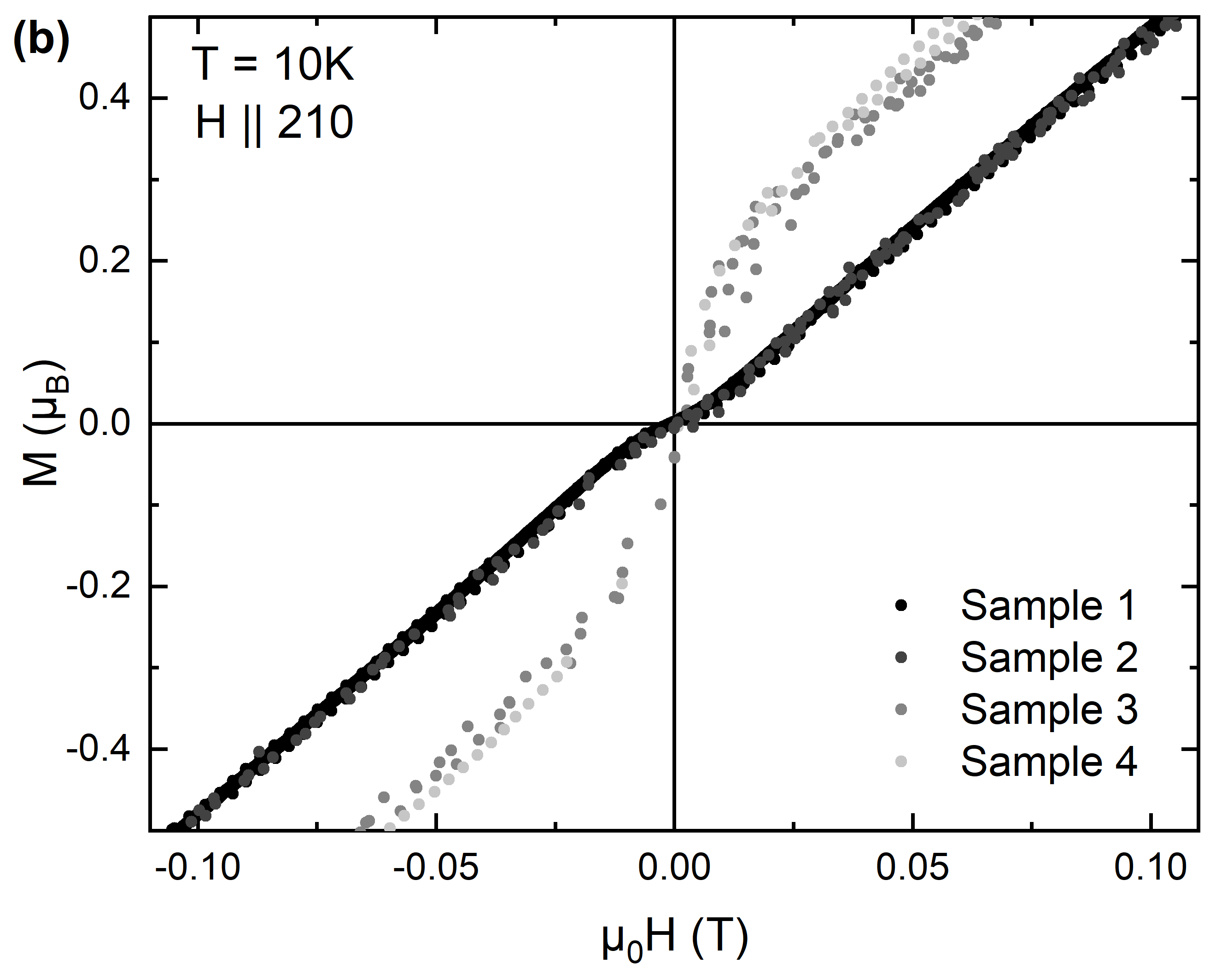}
		\caption{Comparison of (a) the T-dependence of the susceptibility for $\mu_0H=0.01\rm T$ and $\mu_0H=0.1\rm T$ for $H\parallel 210$ and (b) the field dependence of the magnetization at $T=10\,\rm K$ for $H\parallel 210$ for different samples.}
		\label{domains_in_ChiandMvH}
	\end{figure*}

 \vspace*{-\baselineskip}
As shown in Fig.~\ref{domains_in_ChiandMvH}(a,b), the magnetic susceptibility measured at $\mu_0H=0.01\,\rm T$ (green), $\mu_0H=0.1\,\rm T$ (grey) and the field dependence of the magnetization $T=10\,\rm K$ show a sample dependency. The different behavior of the individual samples is probably caused by the presence of different antiferromagnetic domains.

\section{Structural characterization \label{EDX_Laue}}
\begin{figure*}[p]
\includegraphics[width=0.8\textwidth]{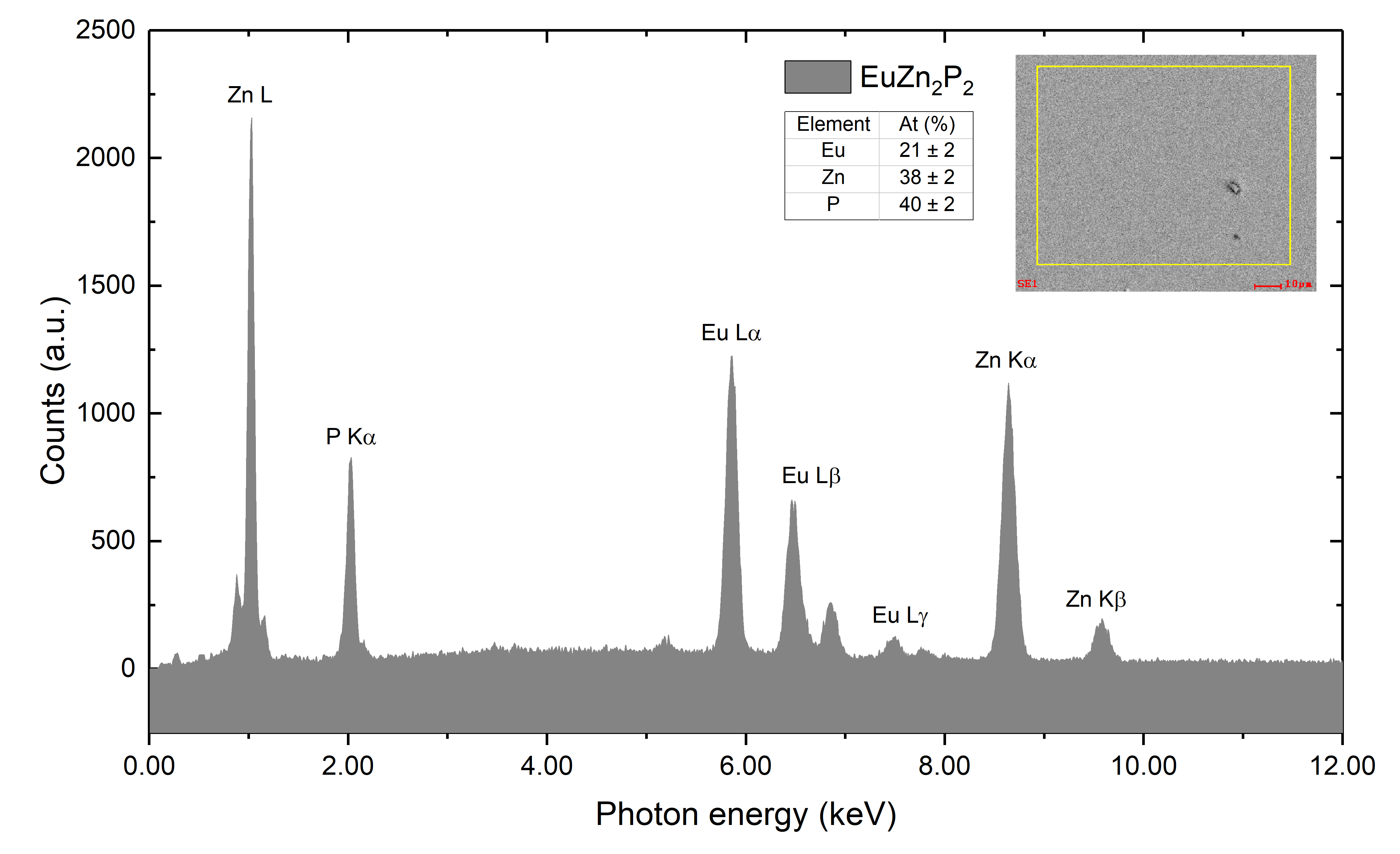}
\caption{Energy dispersive x-ray analysis (EDX) spectrum of the surface of as grown EuZn$_2$P$_2$. The yellow measured area shown in the inset has a size of $~\approx\, (50\times 60)\,\mu\rm m^2$.}
	\label{EDX}
\end{figure*}

\begin{figure*}[p]
  \includegraphics[width=0.8\textwidth]{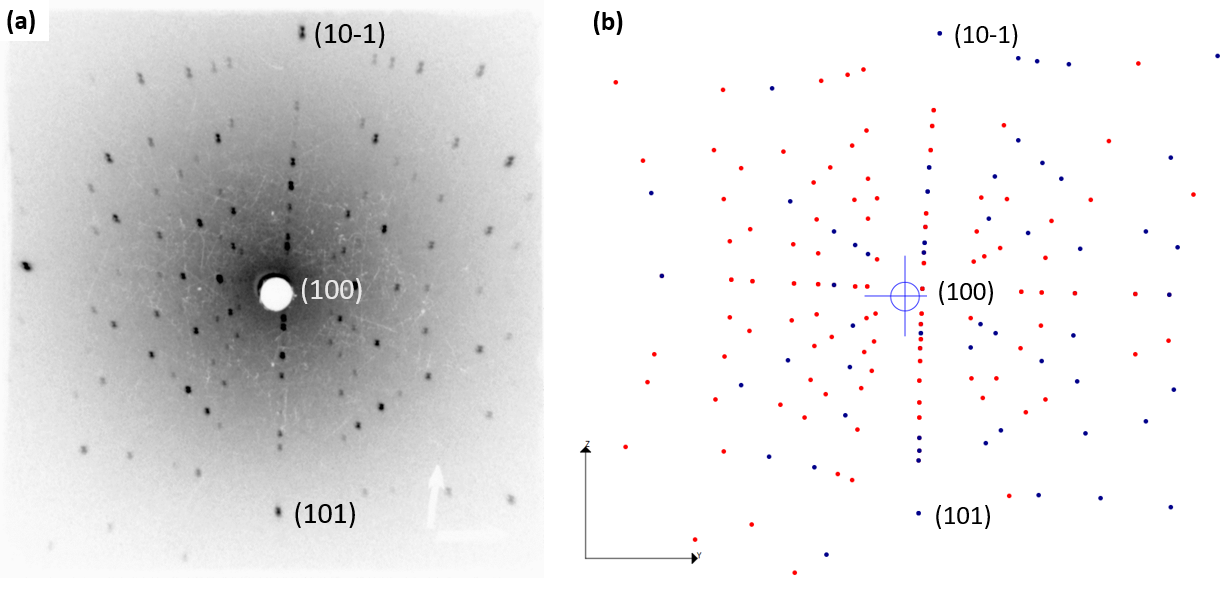}
\caption{(a) Indexed Laue pattern for the beam along (100)-direction in reciprocal space, which is perpendicular to the side edge of the crystal and corresponds to the [210] direction in real space. (b) Laue simulation of the (100) direction in reciprocal space.}
	\label{Laue}
\end{figure*}

 \vspace*{-\baselineskip}
The stoichiometric composition of EuZn$_2$P$_2$ was confirmed using energy dispersive X-ray spectroscopy (see Fig.~\ref{EDX}). The crystal orientation was determined by Laue diffraction. Fig.~\ref{Crystal_Structure}b) shows the Laue pattern recorded with x-ray beam parallel to the crystallographic [001]-direction and Fig.~\ref{Laue} the pattern with beam parallel to the [210]-direction.

\end{document}